\documentclass[12pt]{article}
\usepackage{pdproc}
\usepackage{pdproc,epsfig}
\begin{document}

  \renewcommand{\thefootnote}{\alph{footnote}}

\title{NEUTRINO ASTRONOMY\\AND COSMIC RAYS SPECTROSCOPY\\ AT HORIZONS\footnote{Invited talk at the
III International Workshop NO-VE in Venice,Italy, February 9,2006.} }
\author{Daniele Fargion}
\address{ Physics Department, University La Sapienza,\\Rome, Italy\\{\rm E-mail: daniele.fargion@roma1.infn.it}}
 % \centerline{\footnotesize and}
\abstract{  Air-showering physics may rise in next years at
horizon, offering at different angles and altitudes a fine tuned
and a filtered Cosmic Rays astrophysics and a Neutrino induced
air-showering astronomy; this is possible because of neutrino
masses, their mixing and the consequent replenishment of rarest
tau flavor during neutrino flight into interstellar spaces. Earth
edges and its sharp shadows is a huge detector volume for UHE
neutrino and a noise-free screen mostly for tau air-showers (as
well PeVs anti-neutrino electron air interactions). Satellite
BATSE might, in last decade, already recorded such Upgoing tau
airshowers as Terrestrial Gamma Flashes. ASHRA in Hawaii and CRNTN
in Utah are tracking such fluorescence lights,
 while other projects on Crown array detectors for Cherenkov signals on
mountains, on balloons and satellites are being elaborated and
tested to Tau Air-Showers neutrinos: AUGER, facing the Ande edges,
ARGO located within a deep valley
 may both test inclined showers from the mountains; MILAGRO (and MILAGRITO) on the top
mountain may all  both be triggered by horizontal up-going muon
bundles from the Earth edges; HIRES and better AUGER  detectors,
linking twin telescopes or array scintillators along their
telescope axis may test horizontal Cerenkov blazing photons as
well as nearby Tau airshowers. MAGIC (or Veritas or Shalon)
Telescopes pointing downward to terrestrial ground acts, for EeV
Tau neutrino air-showers astronomy, as a massive tens of $km^3$
water equivalent detector, making it at present the most powerful
dedicated neutrino telescope. MAGIC facing the sea edges must also
reveal mirrored downward UHECR Air-showers Cherenkov flashes.
Magic-crown systems may lead to largest neutrino detectors in near
future. They maybe located on top mountains, on planes or balloons
and in satellite arrays. Amplified Tau-airshower at horizons may
well open a blazing windows, at PeV-EeV energy, to Neutrino
Astronomy. }
  \normalsize\baselineskip=15pt

%%%%%%%%%%%%%%%%%%%% fig 1 %%%%%%%%%%%%%%%%%%%%%%%%%

%%%%%%%%%%%%%%%%%%%%  %%%%%%%%%%%%%%%%%%%%%%%%%
\section{Why Horizontal Air-shower may be disentangled at high altitudes?}
This introductive article offer the minimum of mathematical
equation and the wider view of the title program; I will use the
image communicative  power to describe at least in qualitative way
the multi-face opportunity of the novel theme of horizontal
air-showering from Earth and Space. Indeed while most of low
energy CR ($E_{CR}\sim 10^{9}-10^{12}eV$) are observable from
space, downward higher ones ($E_{CR}\geq 10^{12}-10^{15}eV$) are
deduced by their secondaries in air-showers on Earth ground.
However most of the vertical downward air shower are absorbed at
high altitude ($5\div 10$ km), because their maximal shower slant
depths occurs (for instance at the PeV energy at $X_{max}\simeq
500$g./$cm^2$), in half of the way to ground. Only tiny traces of
muon bundles or Cerenkov flashes are recordable at high altitude
(balloon or mountain) or on array at sea level,\cite{Decor},\cite{NEVOD}, \cite{Cillis2001} \cite{Fargion2004b}. However in the
vertical axis the opening angle and its consequent shower area
($0.1\div1$km$^2$) is quite limited and a few days balloon trip
record time often is unable to catch most rare energetic events.
On the contrary horizontal air-showers observed along the
atmosphere are widely spread by their longer distances ($300\div
500$ km) into widest areas ($10^{1}\div10^{2}$km$^2$). For this
reason observing at different altitude and angle of view the
horizontal air-shower may amplify its area and its intensity
leading to a filter of air-shower primary energy, composition and
cross-section. There is a wide net of shower component to be
correlated  offering, in principle, the possibility to disentangle
the primary energy even by its partial and limited lateral
distribution data. Their arrival angle, the consequent slant
depth, the secondary arrival timing, the electromagnetic
($e^+,e^-,\gamma$) components and their ratios, the muon component
content ($\mu^+,\mu^-$), the net charge presence (due to
geomagnetic shower bending), the pion and nucleon traces
($p,\bar{p},n,\bar{n}$), the Cerenkov signature and the photo
fluorescence signal could trace overall the shower nature and
geometry.
\begin{figure}
\centering
\includegraphics[width=.7\textwidth]{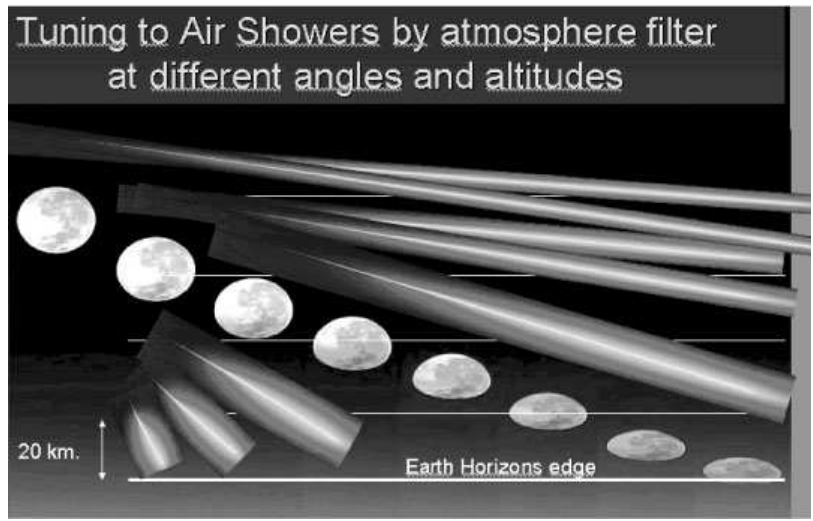}
\caption {The very different High Altitude air-shower morphology
at horizons, whose sizes are tens-hundred time longer, and whose
Cherenkov beaming are thinner (at $35 km$ altitude as small angle
beam as $0.1^o$) than vertical ones ($1.4^o$) because of a much
smaller (up to $0.01$ times respect sea level) air density. Along
some directions (mostly East-West) they are bent and forked in
twin beams by geomagnetic fields. These air-shower profiles (not
in real-scale) are shown for prompt evaluation, in comparison with
the moon size (whose diameter at Earth horizons from Space Station
is as wide as $20$ km.). Also a vertical PeV C.R. air-shower, at
sea level is playing a role of a meter. These air-shower shapes
are shown at right angle to the observer. When beamed in axis to
the observer  their flash may blaze  by Cherenkov lights, either
as an unique or often as a twin (opposite charge) or a triple
(positive-negative neutral components of the shower) light spots.
The twin spots  are as large as tens kilometers each one far from
the other; the separation occurs because of the "polarized"
splitting of the terrestrial magnetic fields.}See
ref.\cite{Fargion2001}\label{fig:fig1}
\end{figure}
%%%%%%%%%%%%%%%%%%%% fig 1 %%%%%%%%%%%%%%%%%%%%%%%%%
 We now remind that vertical showers (as PeV event in
figure) leave only a tiny muon secondary on the ground, less than
$0.1\%$ of the primary energy. The horizontal ones, observed while
$skimming$ the air atmosphere,  offer in principle a much
elongated but rich air-shower, at high altitude ($10-40$ km.), a
narrower jet cone, but a final wider area ($\simeq km^2, 100 \cdot
km^2$) of spread secondaries, contrary to vertical showers. Their
valuable study may calibrate the detector for a more exciting
horizontal air-showers: the up-going ones induced by Neutrino
interactions\cite{Gandhi98} in air or Earth (Tau Air-Showers or Earth Skimming
Neutrinos).
\subsection{The horizontal CR showering spectroscopy: a new view }
 The quantity of information encoded in high altitude
air-shower is so wide that I cannot   summarize it in one single
even introductive paper. There are at each altitude and air
density,a cosmic ray energy and composition in novel condition and
morphology making a new exciting CR showering spectroscopy. For
instance at near a hundred km. altitudes, low energy solar wind
rays may interact by atomic cross-sections lightening auroras in
polar nights; at lower altitudes ($50-60$ km.) low energy
(GeV-TeVs) cosmic rays at horizons are crowding into thin
jet-showers population (tens km length), while at lower altitudes
(near 40 km.) PeV horizontal air-showers are expanding at maximum
size in extreme (few hundred km. long) thin beamed air-showers.
Parallel to them even more rarer and higher energetic C.R. are
showering at maximal length (few hundred km. long) often splitting
by geomagnetic fields into twin forked jets.
 These CR viewing at different quotas could be
imagined, in brief, as a tuning filter for cosmic rays which may
compete and calibrate, above the horizons, with a very different
up-going air showering mostly triggered by
$\bar{\nu_e}+e\rightarrow W^-$ and $\nu_{\tau}+N\rightarrow
\tau+N$, $\bar{\nu_{\tau}}+N\rightarrow \bar{\tau}+N$
interactions \cite{Gandhi98}. More exotic (but allowable even if difficult to
reveal) SUSY ultra-high C.R. traces, energetic neutralinos $\chi$, see\cite{Datta},
may interact with air and electrons making  right-handed s-electron  resonances,
leading to unique electromagnetic air-showers $e +\chi_o \rightarrow \tilde{e_r}\rightarrow e +\chi_o$,
see \cite{Datta}, similar some-how to
Glashow (mostly hadronic) neutrino-electron resonant showers.
These neutrinos air shower at horizons is very competitive with
the conventional muon track $\nu$ astronomy searched in
underground detectors \cite{Anchordoqui}. The main differences between $\mu$ (and
$\tau$) underground (versus air showering lepton) neutrino
telescope is based on four opposite facts:
\begin{enumerate}
  \item Muons are much more penetrating (up to tens PeVs) than tau
  but at EeV regime (and above) the opposite is true: tau exceed
  muon tracks.
  \item High energy Tau might be well revealed  by sampling a few of their unique
  amplified  air shower,  (millions,billions) secondaries, once their up-going
  arrival direction is well recorded. Each downward high energy muon (in $km^3$ detector) should be
  traced for a long path and disentangled by its energy and more over by
   its eventual atmospheric or astrophysics nature \cite{Anchordoqui}.
  \item Tau is unstable and decays in air flight, while $\mu$ is, since
  PeV up EeV energy, extremely stable on any terrestrial size scales. So
  $\mu$ PeV-EeV air shower is not common in air even if partial ($1\%$) atmospheric muon showering
  may shine   above the horizons, possibly to be tested and verified.
  \item Neutrino-induced muons downward are polluted, even in underground
  detector, by million times more atmospheric muons.  Because of the long string
  geometry in AMANDA and most detectors, up-going verticals muons bundles are
  the best view of  water or ice underground arrays.
   Up-going muon above tens TeV (where astrophysical neutrino astronomy should overcome atmospheric neutrino noises)
   are more and more opaque to the Earth diameter. Therefore
  because the Earth opacity the horizons (whose cord are small enough , a few hundred km.) is
  the best direction to search for PeV-EeV neutrinos
  either in underground detectors as well as in air .

\end{enumerate}

 In conclusion above tens-hundreds TeV energy neutrino astronomy may be at discover edge;
 underground muon detector, whose vertical string are viewing at best in vertical axis,
  are polluted (by atmospheric muons) from above, while they are almost blind in vertical
   up-going directions because of the Earth opacity. At horizons
   their geometry make them, up to day, quite an un-efficient detector.
    On the contrary, $\tau$ decay in flight at horizons
     in an extended air-shower is exploiting the unique noise free
     screen (no horizontal air-showers arise beyond the Earth )
       that allow an optimal signals and  easier discover of such amplified UHE
$\nu$ astronomy.

\section{Why  an UHE  $\nu$ astronomy at hand?}
Since Galileo we enjoyed of an optical view first of the planets,
stars,  and later galaxy maps while, since last century  we
enlarged the astronomical electromagnetic windows  in radio,
infrared, UV, X, $\gamma$  with great success. Now a more
compelling UHE $\nu$ astronomy at EeVs energy is waiting at the
corner. It is somehow linked to a very expected new particle
astronomy: the UHECR at GZK energy $\geq 4\cdot10^{19}$eV: it must
be a limited and nearby one (tens Mpc) because of cosmic BBR
opacity. There have been since now two successfully neutrino
astronomy at opposite low energy windows: the solar and the
supernova ones. The solar ones has been explained by Davis,Gallex,
SK, SNO experiment in last four decades opening the $\nu$ physics
to a solar neutrino mass splitting and a clear probe to its mixing
behavior. The supernova SN 1987A was an unique event that anyway
had a particular expected signatures at tens MeV. On going
experiment on cosmic supernova background in S.K. are at the
threshold edges, possibly ready to a
 discover of this cosmic background. However there is a more exciting and
energetic $\nu$ astronomy at PeV and EeV energy associate to the
evidence of charged UHECR spectra at EeV and tens hundred of EeV
band. Indeed any EeV CR originated nearby an AGN or GRB or BL Lac
jet will be partially screened by the same source lights leading
to a consequent photo-pion production, associated with PeV
secondary neutrinos. In a much simpler and guaranteed way, at
energy about $4\cdot10^{19}$eV, UHECR should propagate in cosmic
photon black body, being partially arrested  by  photopion
productions, (GZK cut-off), leading to EeV neutrinos all along the
Universe confines. These UHE $\nu$ components, consequence of the
GZK cut-off, are called cosmogenic or GZK neutrinos. Their flux
may be estimated by general arguments and there is quite a wide
consensus on such neutrino GZK flux at EeV energies. These
\textit{guaranteed} neutrinos may be complementary to possible
\textit{expected} higher energy neutrinos (at ZeV energies) whose
role might explain UHECR isotropy and homogeneity being originated
at cosmic distances. In this model UHECR   born as nucleons via
$\nu+\bar{\nu_R}\rightarrow Z\rightarrow X+N$ (Z-burst or Z-shower
model) \cite{Fargion-Mele-Salis99},\cite{Weiler97},\cite{Yoshida1998}
are overcoming present (AGASA,HIRES,AUGER) un-observed
local (VIRGO,PERSEUS) source distribution, as would be prescribed
by naive GZK cut-off. However GZK neutrinos and Z-Burst neutrinos
at EeV are making comparable flux predictions and we shall
restrict to the simplest GZK flux assumption.

\subsection{Six Neutrinos in search of an Author and an  Astronomy: historical connections}

%%%%%%%%%%%%%%%%%%%% fig 3 %%%%%%%%%%%%%%%%%%%%%%%%%
\begin{figure}
\centering
\includegraphics[width=.7\textwidth]{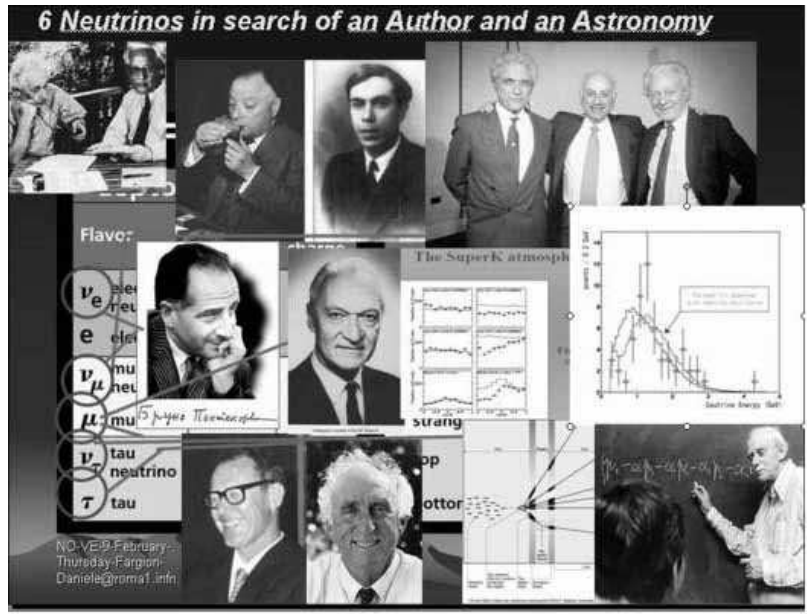}
\caption {The Six main actors (Pauli,Majorana, Pontecorvo,
Conversi, Steinberger, Perl) in Lepton search among the great
corner shadows of Dirac and Einstein-Bose (Fermions-Bosons);
B.Rossi (cosmic rays) and SK , SNO, K2K achievements at the center
remind us of the cosmic rays neutrinos, their mixing and the new
tau neutrino role in air-showers. }\label{fig:fig1}
\end{figure}
%%%%%%%%%%%%%%%%%%%% fig 3 %%%%%%%%%%%%%%%%%%%%%%%%%
The consequent three flavour matter-anti-matter neutrino states,
\textit{$3\times 2 =6$ neutrino actors in search of an author} are
linked to a wide historical sequences of nomes: Pauli (by energy
conservation law), Majorana (because of a possible
matter-antimatter overlapping), Pontecorvo (for neutrino
oscillations among themselves or among different flavour nature);
indeed a main actor that opened the key question for neutrino
mixing and multi-flavour leptons is the muon (Who ordered that?):
its discover was made by Conversi,(Pacini,Piccioni) on 1948
opening the road to much later and actual lepton and quark family
frame (clarified by Gell'mann, Neeman, Cabibbo, Glashow, Salam,
Weinberg, since last $40$ years). The muon neutrino discover is
linked to Leon Lederman, Melvin Schwartz and Jack Steinberger,
shown on the right side picture. The fundamental discovers in
cosmic ray, disentangling their muon and pion composition, has
been inspired by a long list of revolution and instrumentations
invented by Bruno Rossi (in the central picture). His main role in
CR history was fundamental for understanding the complex
air-shower and in opening X-gamma astronomy; his results are the
backbone of most paper in this field. The additional actor on
neutrino frame is the more recent Perl's discover in 1970 of a
third unexpected lepton, the tau, whose associated neutrino has
been experienced only thirty years later on this century. Many
more discovers by Davis, Gallex, S.K., SNO, K2K, are testing the
earliest B.Pontecorvo predictions and their cosmic roles.

%%%%%%%%%%%%%%%%%%%% fig 1 %%%%%%%%%%%%%%%%%%%%%%%%%

%%%%%%%%%%%%%%%%%%%% fig 1 %%%%%%%%%%%%%%%%%%%%%%%%%
%%%%%%%%%%%%%%%%%%%%%%%%%%%%%%%%%%%%%%%%%%%%%%%%%%%%%%%%%%%%

All over those three light neutrino flavours (verified also at LEP
on 1990s by the Z boson width) the role of Dirac neutrino is
competitive to Majorana one and doubling neutrino states are
possible.The  recent definitive discover of atmospheric
neutrino \textit{anomaly} on 1997,  by SuperKamiokande and last
 SNO verification of solar neutrino nature are leading to a
first precise picture of two neutrino mass splitting narrowing
flavour mixing areas.The atmospheric neutrino \textit{anomaly}
(1997) guaranteed the birth of tau neutrino astronomy as was
immediately noticed \cite{Fargion 2002a}.To conclude the
historical picture one must remind the solar understanding by
Bachall and SNO, the fermion electro-weak understanding by Fermi
first and the Weinberg-Glashow-Salam Standard model, later ,
confirmed and revealed by C.Rubbia last decade.

\section{Why upward tau air showers are linked to neutrino mass and
mixing?} The tau production is limited,in general,to high energy
charmed mesons,whose productions are rare and severely suppressed
respect to lower energy pions ones. For this reason
$\nu_{\mu}$,$\bar{\nu_{\mu}}$ astronomy had a major attention in
last century, also for the deeper $\mu^+$,$\mu^-$ penetration with
respect $e^+$,$e^-$ and unstable tau. However the definite
$\nu_{\mu}\leftrightarrow\nu_{\tau}$ (SK data) disappearance and
the flavour neutrino mixing has given to
$\nu_{\tau}$,$\bar{\nu_{\tau}}$ a new life and attention. Indeed
the additional possibility to oscillate, even at highest energy
($10^{19}$eV) energy and  lowest mass splitting ($\Delta
m\simeq10^{-2}$eV),  is guaranteed by the huge stellar galactic
and cosmic distance ($\gg$hundred pc).
\begin{equation}\label{distanza}
  L_{\nu_{\mu}\rightarrow \nu_{\tau}}=8.3pc\biggl(\frac{E_{\nu}}{10^{19}eV}\biggr)
  \biggl({\frac{\Delta m_{ij}^2}{10^{-2}eV ^{2}}}\biggr)^{-1}
\end{equation}
respect to the above oscillatory one. In some sense this $\tau$
neutrino astronomy offers additional proof of $\nu$ mixing.It
should be noticed that on principle $\nu_{\mu}\rightarrow
\nu_{\tau}$ appearance may (or is going to) be revealed in SK
events.However the conjure of $\tau$ large threshold energy (4GeV)
and the small Earth radius size make this possibility a different
or marginal one.On the contrary a solar flare neutrino $\nu_{\mu}$
may travel and reach the Earth at threshold tens GeV energy and
convert itself successfully into $\tau$ leading to a possible
$\nu_{\tau}$ neutrino astronomy from solar flare \cite{FM}.

\subsection{Neutrino Showering in Universe: Why a Z-Burst solves GZK puzzle?}

%%%%%%%%%%%%%%%%%%%% fig 4 %%%%%%%%%%%%%%%%%%%%%%%%%
\begin{figure}
\centering
\includegraphics[width=.4\textwidth]{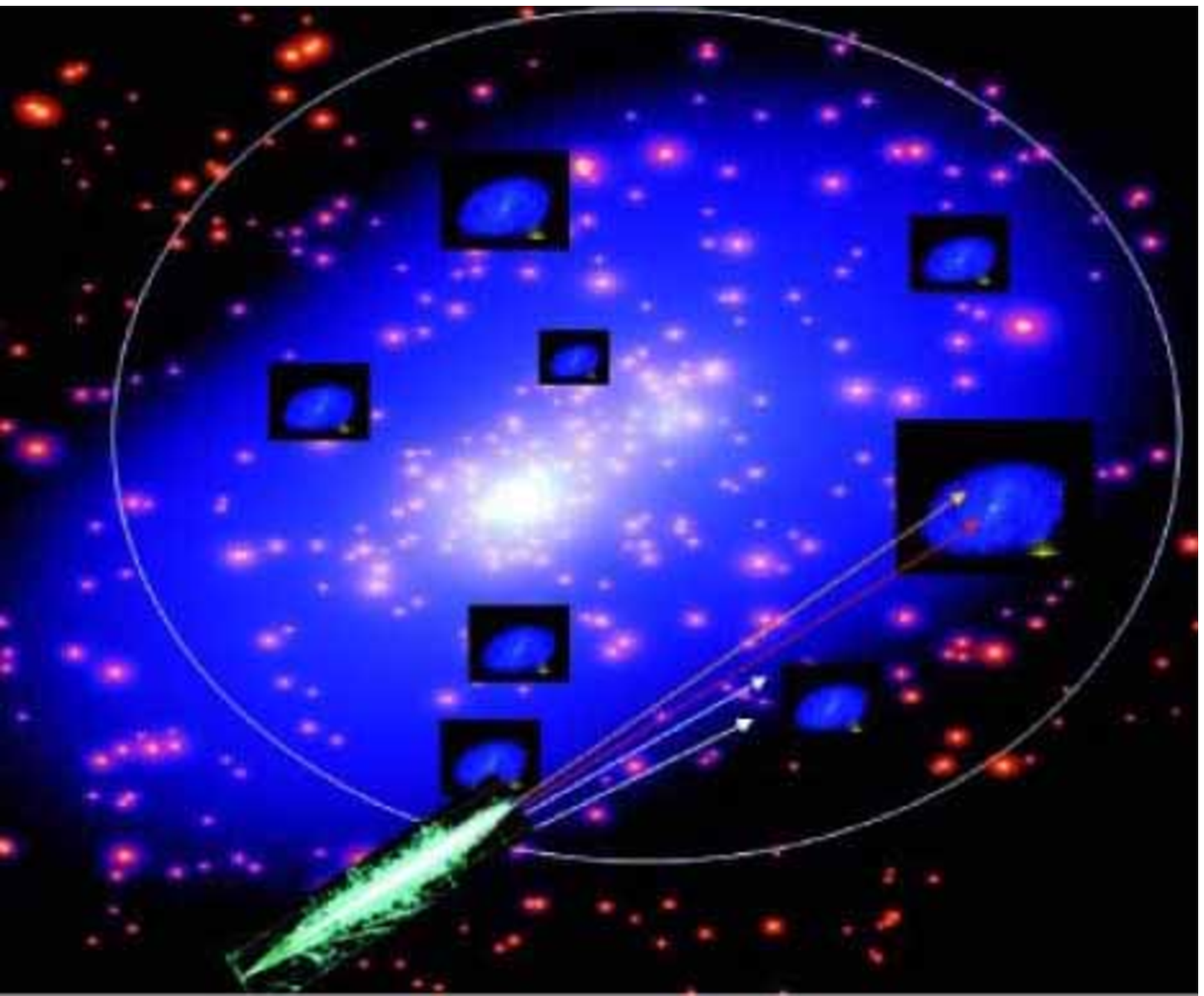}
\caption {A schematic cosmic map where ZeVs neutrinos from BL Lac
at cosmic edges, interact on relic one leading to Z-Burst and
showering into UHECR. The white ellipse (a ten of Mpc. size)
contains most of a super-galactic cluster where a relic light
($0.4-0,1$ eV.) neutrino clouds are smeared as a diffused hot dark
matter component. The Z-boson decay contains UHE neutrons (white
color secondary arrows) and proton and anti-proton longer life
UHECR. An analogous (but more exotic process takes place for UHE neutralino scattering onto relic neutrino
leading, via sneutrinos resonances, to UHE showering in space. Because of the
higher mass of sneutrinos are generated at higher (than neutrino), tens -hundred ZeV neutralino energies
 }.See \cite{Fargion-Mele-Salis99},\cite{Datta}.\label{fig:fig1}

\includegraphics[width=.4\textwidth]{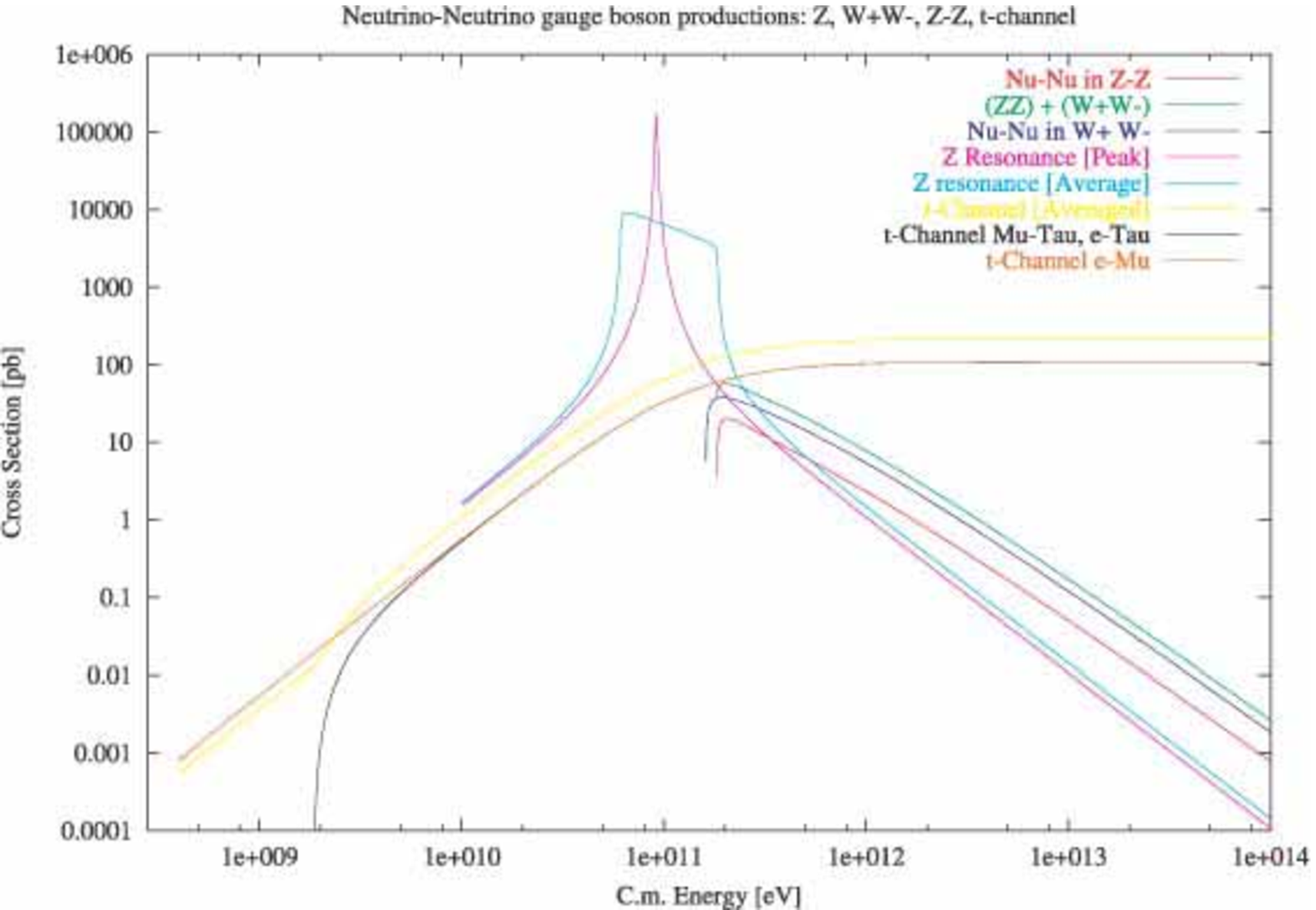}
\caption {The cross-section for UHE neutrino-anti-neutrino
s-channel leading to resonant Z-Boson decay; additional WW and ZZ
channels are also shown. The possible mild relativistic relic
neutrino distribution will induce a Doppler shift whose smearing
role will lead to a "broken-tower" peak shown in figure. The
consequent Z-Burst shower will feed electro-magnetic (electron
pairs,gamma)and nuclear components; the latter may be source of
UHECR escaping GZK cut-off.}See \cite{Fargion-Mele-Salis99},\cite{Fargion2000}\label{fig:fig1}

\includegraphics[width=.5\textwidth]{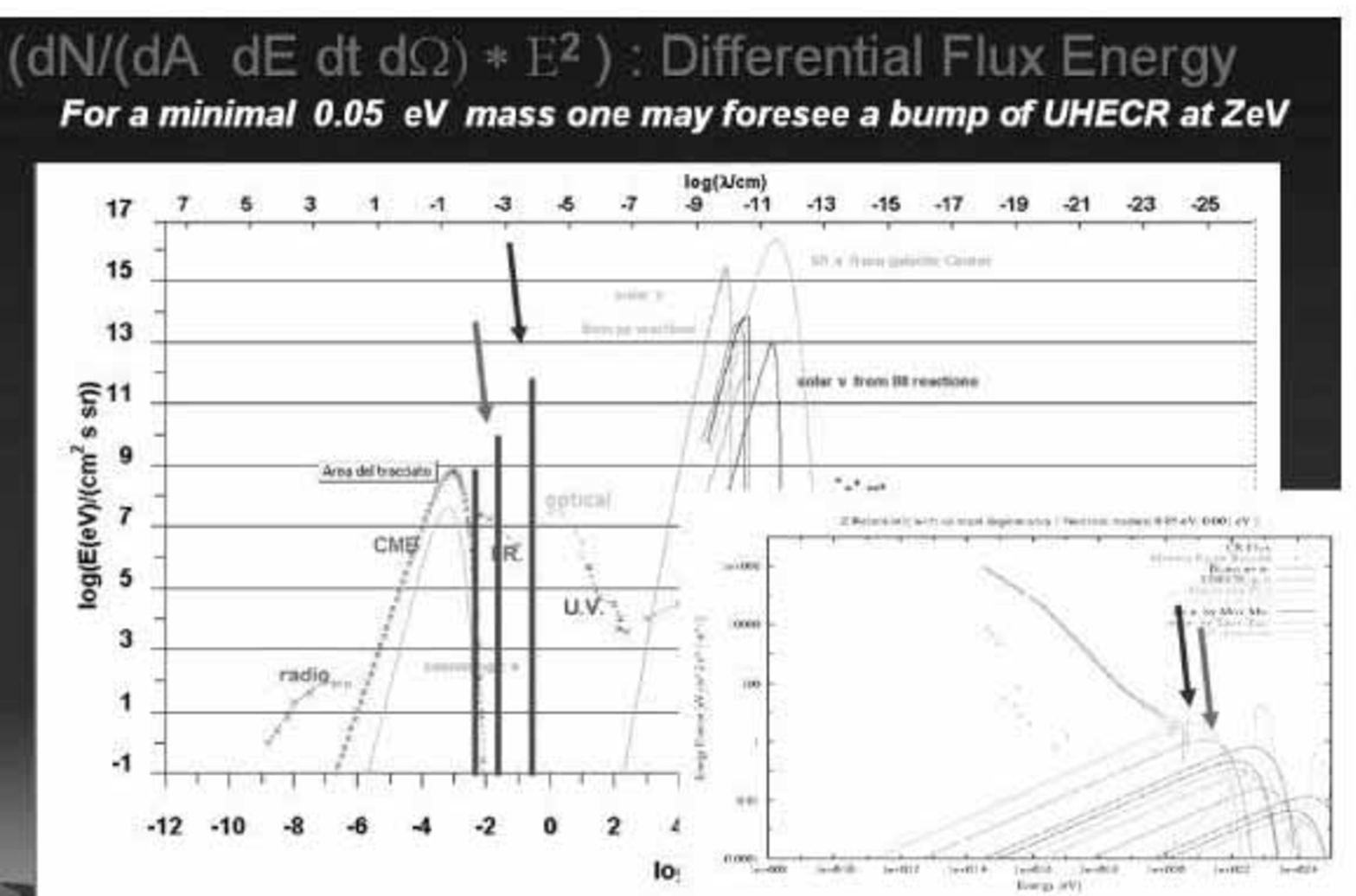}
\caption {The energy fluence in radiation and particle in our
Universe and by solar neutrino influence. The relic neutrino
masses induce a possible dominant component at different (almost
equal or non-degenerated) eV masses. Their presence will rule a
Z-Burst showering and an UHECR tails just at the edge of the
cosmic ray scale. Similar showering maybe born by UHE ZeVs
neutralino scattering into relic neutrinos via sneutrino
resonances . The lighter the neutrino mass the
 higher its UHECR influence, up to ZeV UHECR edges.}See \cite{Fargion-Mele-Salis99},\cite{Fargion2000},\label{fig:fig1}
\end{figure}
%%%%%%%%%%%%%%%%%%%% fig 6 %%%%%%%%%%%%%%%%%%%%%%%%%

As a cosmic ray hitting atmosphere,in analogous way may shower an
UHE $\nu$ hitting a relic $\bar{\nu}$:in hot and spread dark
halo,may lead to Z boson and its decay. Among the UHE secondaries
nucleons (and anti nucleons) one may reach the Earth appearing as
an UHECR. Because the relic neutrinos $\nu_{r}$ mass may be about
or below $0.4$ eV the incoming UHE $\nu_{i}$ should be above
$E_{\nu_{i}}\geq M_Z^2/m_{\nu_{i}}\simeq 10^{22}\cdot
(m_{\nu_{i}}/0.4eV)^{-1}$ eV, while the final nucleon shares an
average $E_p \simeq 2\cdot 10^{20}eV \cdot
(m_{\nu_{i}}/0.4eV)^{-1}$,in agreement with AGASA data. The UHE
ZeV neutrinos may escape BBR opacity and connect UHECR to cosmic
sources and explain their cosmic homogeneous and isotropic imprint
(as most observed UHECR maps show). While recent report from Hires
 and AUGER seem to disclaim the AGASA spectra absence of a cut, the missing  nearby
 UHECR sources correlated with these UHECR events leaves (in my opinion) still open the GZK puzzle.
 The presence of non degenerated lightest neutrino mass
($m_{\nu}\leq 0.1 eV$) may offer the presence of multiple
Z-resonant UHE neutrino energies corresponding for example to
$E_{\nu}=4\cdot10^{22}(m_{\nu}/0.1eV)^{-1}$ eV or
$E_{\nu}=8\cdot10^{22}(m_{\nu}/0.05eV)^{-1}$eV. If this will be
the case we must $foresee$ for such a lightest case , a future
highest energy bumps in UHECR : indeed the possible maximal
atmospheric neutrino mass (assuming negligible all other neutrino
masses), $m_i=\sqrt{\Delta m^2_{atm}}\simeq 0.05eV$, implies
ultimately a limiting resonance at
$E_{\nu}=8\cdot10^{22}(m_{\nu}/0.05eV)^{-1}$ eV, whose consequence
may reflect in an unexpected energy injection and un upper limit
UHECR bump at $E_{nucleon}=1.6(m_{\nu}/0.05eV)^{-1}$ ZeV energy
edge. This bump maybe mitigated by lower neutrino clustering
density, but also increased by little larger cosmic (GZK)
distances. The AUGER and the EUSO detectors may reach these
extreme goals possibly discovering not a \textit{neutrino spectra
deeps} \cite{Quigg} but the more concrete imprint of the $emergence$
in the UHECR spectra edges of relic $\nu$ masses shadows. It
should be noted that the even extreme light relic $\nu$ masses
($0.1$eV) may be spread inside cosmic volumes whose radiuses may
be comparable with GZK cut-off. This possibility implies an
efficient GZK suppression of Z-Burst secondaries gamma
component,at $10^{19}$eV, because they are much more absorbed than
nuclear UHECR component at this energy. This peculiarity may
reconcile the apparent absence of $\gamma$ UHECR at $10^{19}$eV
signals, while being still consistent  to most Z-burst model
predictions,\cite{Fargion-Mele-Salis99}. Let us remind that a possible very light
relic $\nu$ at nearly relativistic regime should  spread the
corresponding $\nu\bar{\nu}$ Z resonance peaks leading to a
smoother bump whose "broken tower"  shape has been first
recognized \cite{Fargion2000} and  recently re-discovered by many others
authors.

\section{Why Air showering by W$^-$ resonance and  $\nu_{\tau} \rightarrow {\tau}$, $\bar{\nu_{\tau}} \rightarrow \bar{{\tau}}$, in air?}

As the Z boson peak  favors UHE neutrinos in Z-shower model for
light neutrino masses ($E_{\bar{\nu_{e}}}\simeq
m_Z^2/2m_{\nu}\simeq 10 $ ZeV $\frac{0.4 eV}{m_{\nu}}$), in the
same way $\bar{\nu_{e}}e\rightarrow W^-$ resonance
($E_{\bar{\nu_{e}}}\simeq m_W^2/2m_e\simeq 6.3 PeV$) favors
energetic $E_{\bar{{\nu}_e}}$ hitting and showering beyond
mountain barrier (as well as within air horizontal edges). The
advantage of a mountain lay is double:a sharp filter for all the
horizontal hadronic air shower (and even muon tails) and a dense
beam dump where $\bar{\nu_{e}}e\rightarrow W^-$ or
$\nu_{\tau},(\bar{\nu_{\tau}})+N\rightarrow \tau (\bar{\tau})+X$
may take place.These events are double (first $\nu N$ or
$\bar{\nu_{e}}e$ event and later a $\tau$ decay);in water the
phenomenon has been noticed nearly ten years ago , see \cite{Learned Pakvasa 1995}.
The idea of this $\tau$ showering in water was been
considered as the double bang signatures, rarely observable in
km$^3$ detector. The some double bang reformulated \textit{in} and
\textit{out}, \textit{in} rock mountains (or Earth) first and \textit{out}
within air,later, was the main proposal  discussed first  at
the end of the previous century \cite{Fargion1999}. In particulary since  six years
ago , see \cite{Fargion 2002a}, the upgoing and horizontal air showers has been widely
formulated for detention beyond mountain chain, as the Alps and Ande ones. These ideas had
been promptly considered for on going AUGER experiment, just
nearby Ande mountain chain, see \cite{Fargion 2002a}\cite{Fargion2004},\cite{Miele et. all05}.
 Later on the same idea of \textit{old}
and \textit{regenerated} horizontal air shower have been
considered by other authors \cite{Bertou2002} as well as by a wide list of
additional authors \cite{Feng2002},\cite{Tseng03}, \cite{Jones04},\cite{Yoshida2004}. The difference of the $\tau$ role in its
crossing the Earth lay is in its complex energy loss processes.
Ionization, bremsstralhung, pair production and photo-nuclear
losses are suppressing the $\tau$ primary energy in such way its
own lifetime may be shortened  suppressing its propagation length.
The understanding of the correct interaction length has been
noticed by \cite{Fargion1999} and it has been incorporated on 2000 by \cite{Fargion 2002a},
and in the complete final $\tau$ radiation length \cite{Fargion 2002a},\cite{Fargion2004}.
While first and late attempts assumed a fixed $\beta$ parameter or
a linear ones, the $\beta$ dependence with its logarithmic growth
with energy has been considered correctly by \cite{Fargion 2002a} (and not other authors) as
it has been  probed
in detail only recently \cite{D05}.

\section{How penetrating is a $\tau$ length versus muon one?}
One of the most common place in $\nu$ telescope astronomy is to
consider $\mu$ because more penetrating than $e$ and $\tau$ \cite{Fargion 2002a}. This
is true in the TeV-PeV energy. However the PeV $\tau$ is already
 to escape a mountain, decay in
flight and amplify its shower loudly, respect to a single $\mu$
escaping at some energy from a mountain. Moreover the muon
logarithmic growth  is reached at EeVs by a linear growth of an
UHE $\tau$ , mostly because if the lepton is heavier, its
electromagnetic loss is smaller. Un-fortunately hadronic losses do
not allow the $\tau$ to increase its penetration but $\tau$ is
more penetrating a those UHE regime where $\nu$ astronomy overcome
the atmosphere $\nu$ noise, see \cite{Fargion 2002a}. In more sophisticated approach, not
shown here for sake of simplicity, one may estimate the Earth skin
to Tau Air-shower for $shorter$ maximal lengths that guarantee a
unsuppressed $highest$  Tau escape energy; this minimal Earth skin define a
smaller volume and lower tau air-showering rate, but at highest
$EeVs$ energy \cite{Fargion2004}.

%%%%%%%%%%%%%%%%%%%% fig 7 %%%%%%%%%%%%%%%%%%%%%%%%%
\begin{figure}
\centering
\includegraphics[width=.6\textwidth]{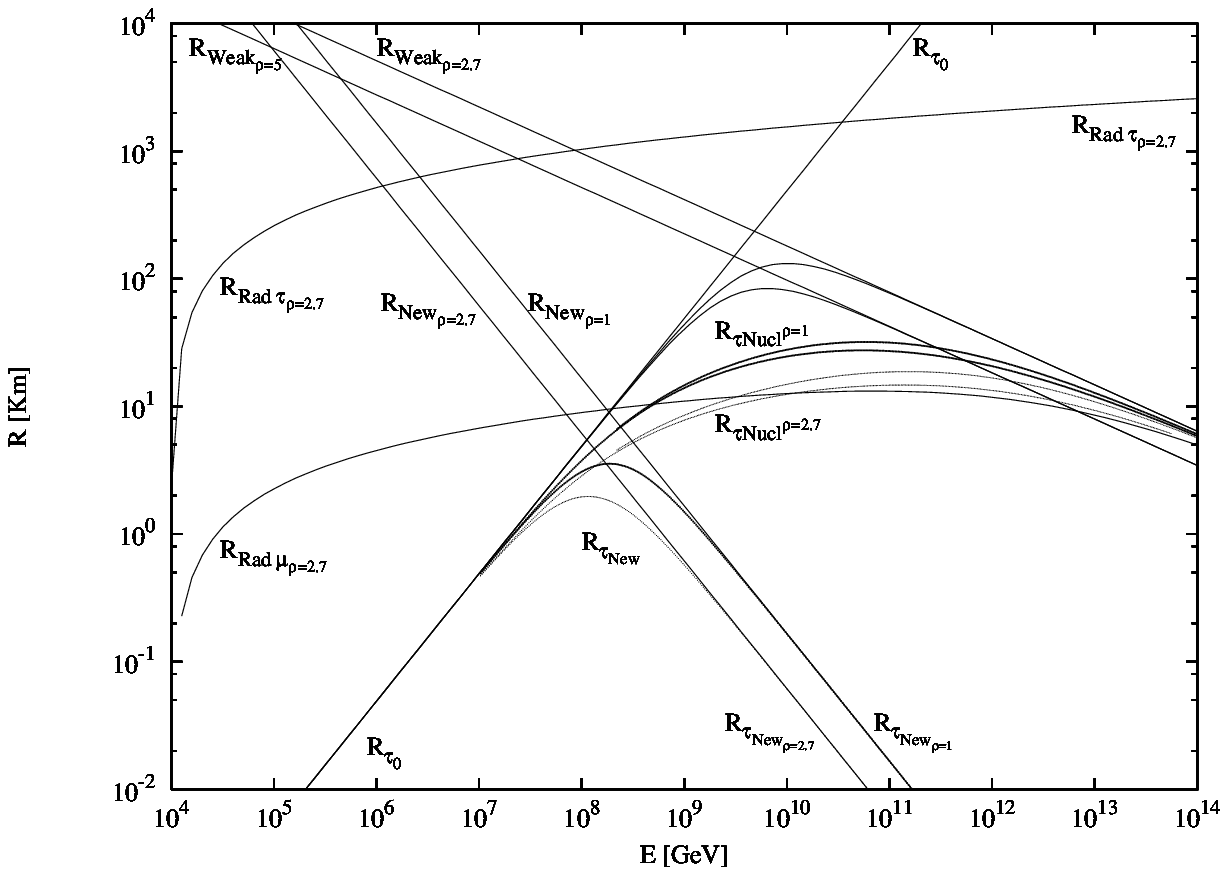}
\caption {While tau are extremely unstable and  electron are
leading to short radiation length, muons are usually the most
penetrating lepton; this usually favor underground muon neutrino
telescope.However at ultra-relativistic regime tau reach and
overcome the muon tracks, making the heaviest lepton the most
penetrating. Because life-time linkage to tau energy and to its
energy losses,dominated by hadron and pair production, the tau
interaction length is derived by an hybrid transcendent equation
comparing energy losses and life-time length.  }\label{fig:fig1}
\includegraphics[width=.7\textwidth]{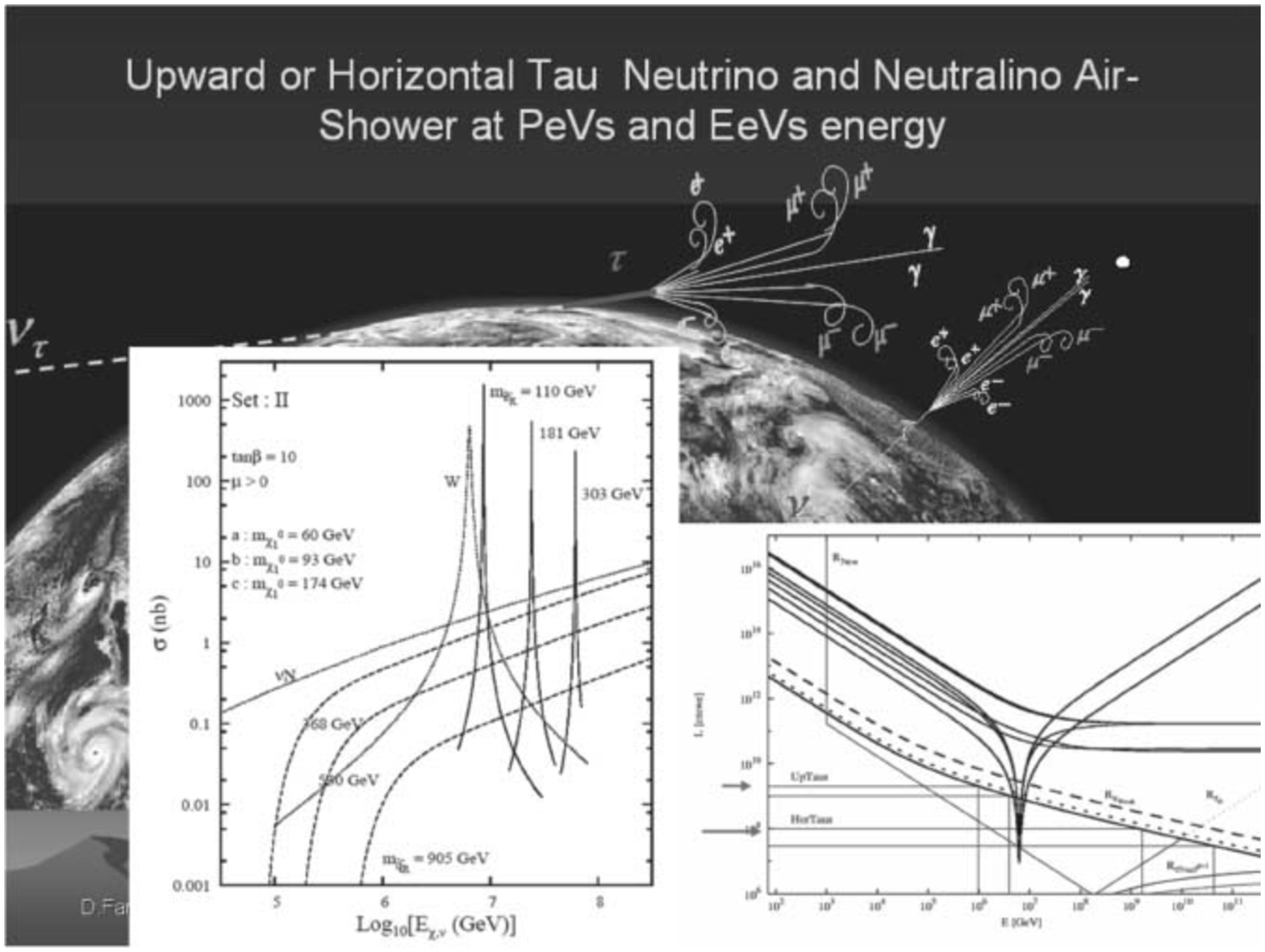}
\caption {A schematic view of the possible Horizontal Tau
Air-showers at EeVs energy versus a lower PeV vertical up-going
Tau Air-Shower. In the left side insert the cross-sections for UHE anti-neutrino
with electrons, mediated  by $W^-$ and the almost comparable UHE neutralino scattering on electron
leading to $\tilde{e_R}$  whose decay in flight lead also to UHE electromagnetic jet-shower.
The consequent interaction length for
both neutrino with nucleons and its peculiar
anti-neutrino-electron  Glashow resonance is shown in the second insert.
 The Earth diameter in nearly $10^{10}$ cm.
water equivalent; therefore  the terrestrial neutrino opacity
arises above tens PeV energy or inside the narrow  resonant
Glashow's neutrino peak. To overcome this neutrino opacity one may
consider mountain chains or small (PeV) Uptaus or shorter
terrestrial cord, for higher energy Hortaus,that are just at the
horizons as shown by the red arrows } see \cite{Datta}.\label{fig:fig1}
\end{figure}
%%%%%%%%%%%%%%%%%%%% fig 2 %%%%%%%%%%%%%%%%%%%%%%%%%

%%%%%%%%%%%%%%%%%%%% fig 2 %%%%%%%%%%%%%%%%%%%%%%%%%
\begin{figure}
\centering
\includegraphics[width=.7\textwidth]{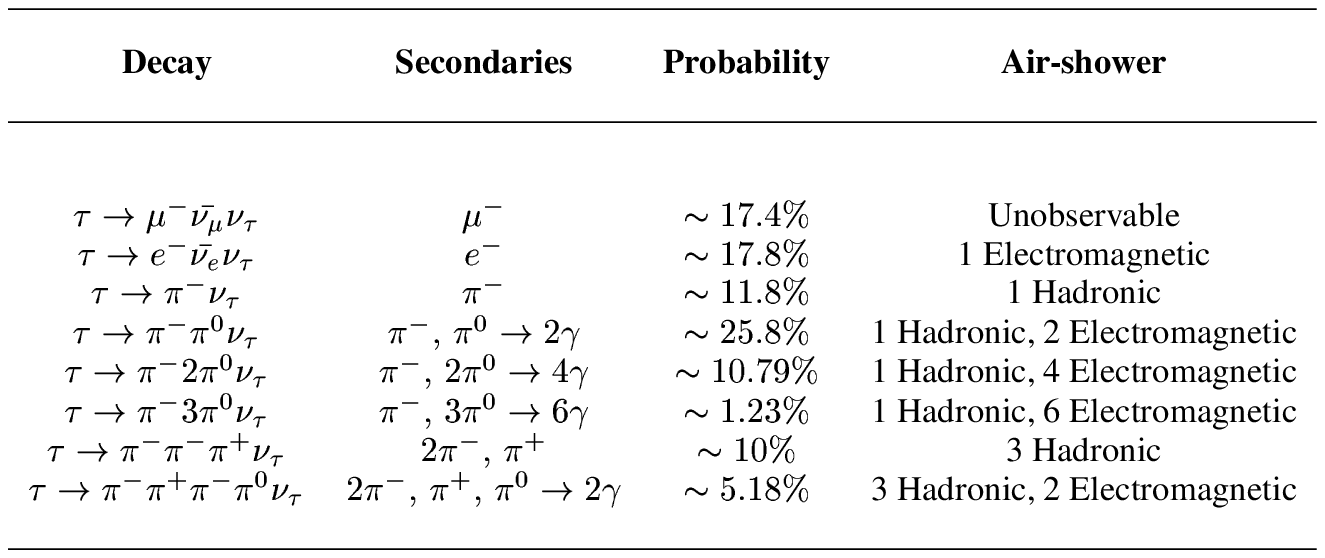}
\caption {The possible Tau decay channel and corresponding
air-showering mode  }\label{fig:fig1}
\end{figure}
%%%%%%%%%%%%%%%%%%%% fig 2 %%%%%%%%%%%%%%%%%%%%%%%%%
The behavior of $\tau$ lengths for most adopted energy losses,the
possible definition of a shorter length that guarantee a higher
outgoing $\tau$ energy,all the detailed Earth profile density for
escaping $\tau$ and the consideration of the finite atmosphere
size for escaping $\tau$ air shower, all these details have been
analyzed in a tail of recent article \cite{F04}.Independent
attention has grown in studying the upgoing $\tau$ flux in km$^3$
detectors \cite{Feng2002},\cite{Tseng03}, \cite{Jones04},\cite{Yoshida2004}.
The general results are not always
converging and a summary of the most recently results has been
shown (see last fig  in ref.\cite{Fargion2004} for general comparision).
\subsection{The table of PDG $\tau$ air shower}
 While we have not yet experienced $\tau$ definitive
air showers , we may foresee that any neutrino $\tau$ astronomy
will, soon or later, test the $\tau$ decay channels. Indeed the
main multiple $\tau$ decay channel are leading to weighted channel
and showers described in PDG table. It will be possible,in
principle,to verify by ratio of $\bar{\nu_e}e\rightarrow
W\rightarrow \tau$ \textit{monocromatic} channel versus
$\nu_{\tau}+N\rightarrow \tau$ channel, the $\nu_{\tau}/\nu_e$
abundance and the primary flavour mixing. In a few words $\tau$
air shower must be consistent and correlated, in its decay mode by
electromagnetic, hadronic and hybrid channel, with well known
elementary particle result.

\section{Eleven Present, Past and Future Experiment in search of a Tau}
  There are very advanced experiment that ( wherever they are aware or not) might
  point for  Tau Air-Showers, even they were originally thought for
  other scientific targets.  The list of these High energy Showering experiment
     adaptable to tau Neutrino and Horizontal Showering is here briefly reviewed:
    1)Argo, in Tibet; 2) Milagro (and Milagrito) in USA mountains,3) AUGER experiment in Argentina, 4) Space
    Station Crown Arrays (to be effectively proposed \cite{FarCrown}), 5) EUSO telescope, 6)  BATSE satellite in CGRO ($1991-2000$ past),
    7) ASHRA experiment in Hawaii, 8) CRTNT Fluorescence array in Utah or China,\cite{Cao} 9)
    Muon Array Telescope in Jungfraujoch,\cite{Iori04} 10)  Cherenkov
    Telescopes on High Mountains facing the mountains, like Shalon in Kazakistan.
    In this view the Magic Stereo
     (as well as Veritas array) Telescopes facing the Earth edges are somehow ideal \cite{Fargion2005}.
     The first $Horizons$ tests maybe done possibly in cloudy and otherwise astronomical useless
     nights.   Here below the images and the
  captions explaining how those experiments may find Tau Air-Showering by a minimal optimized
  trigger set up.

  \subsection{Argo}
  This large area array inside  a deep valley in Tibet may record
  PeVs Tau air-showers emerging from the mountains around. The
  nearby Chines-Japanese twin experiment may enlarge the area. The
  presence of more (tens-hundreds) spread (small, few $m^2$ area) elements at hundred
  meters one from the other, in vertical structure as well as the
  covering of the inner wall periphery of the detector house, may
  greatly increase the ARGO ability to reveal PeVs air-showering
  below the mountain shapes. The  variable opacity to atmospheric
  GeVs-TeVs muons within the mountain shadows,  is a needed test.
  %%%%%%%%%%%%%%%%%%%% fig 8 %%%%%%%%%%%%%%%%%%%%%%%%%
\begin{figure}
\centering
\includegraphics[width=.6\textwidth]{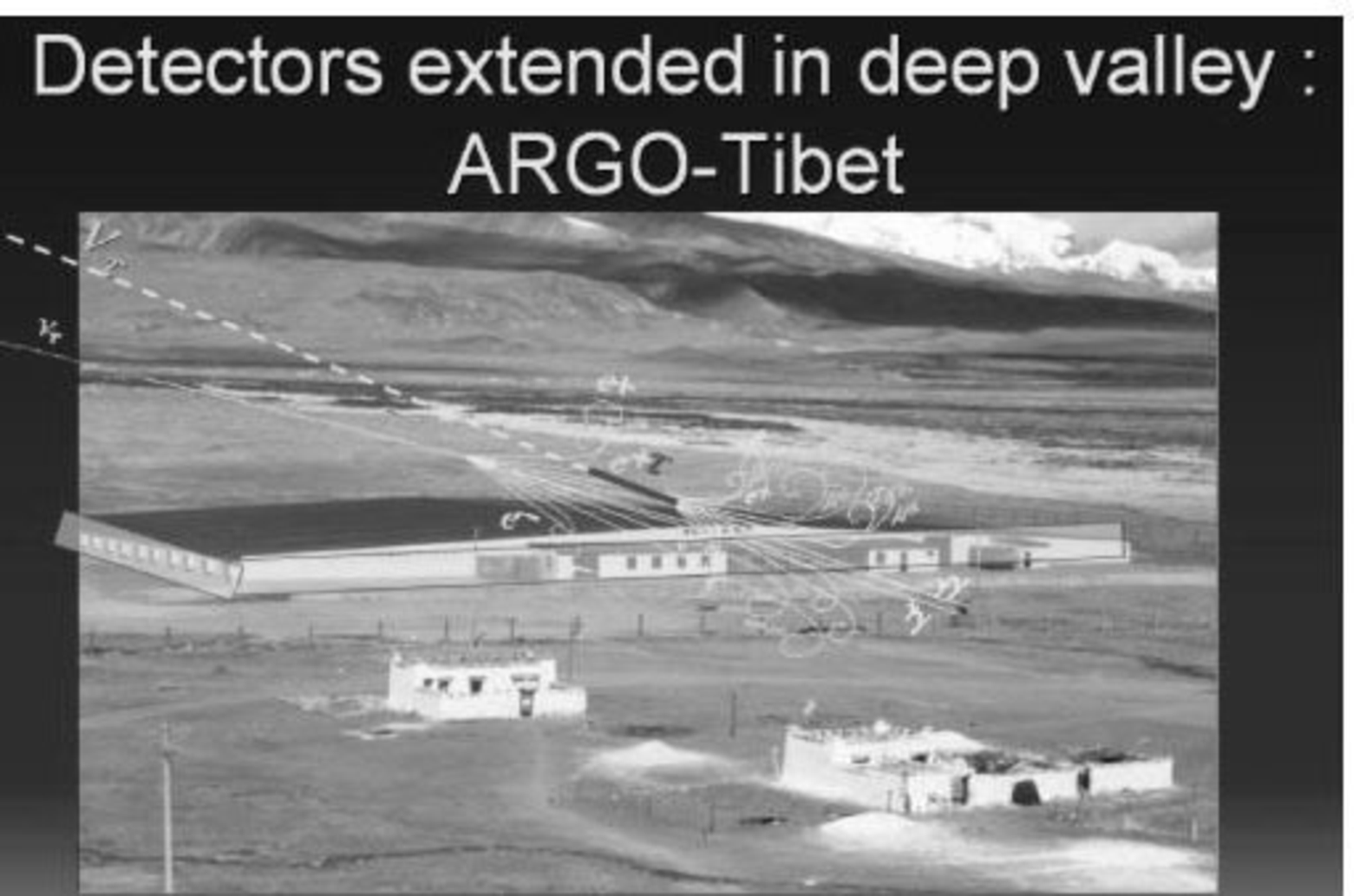}
\caption {The possible use of Italian-Chinese ARGO (and its twin
nearby Chinese-Japanese array) to monitor, inside a wide deep
valley, inclined or horizontal tau air-showers originated by
surrounding mountains; the signal may be better revealed by
additional array detectors on the walls along the lateral
boundaries; these lateral-wall array are  in analogy to  present
Nevod-Decor detector parallelepiped  structures, in
Russia.}\label{fig:fig1}

\includegraphics[width=.6\textwidth]{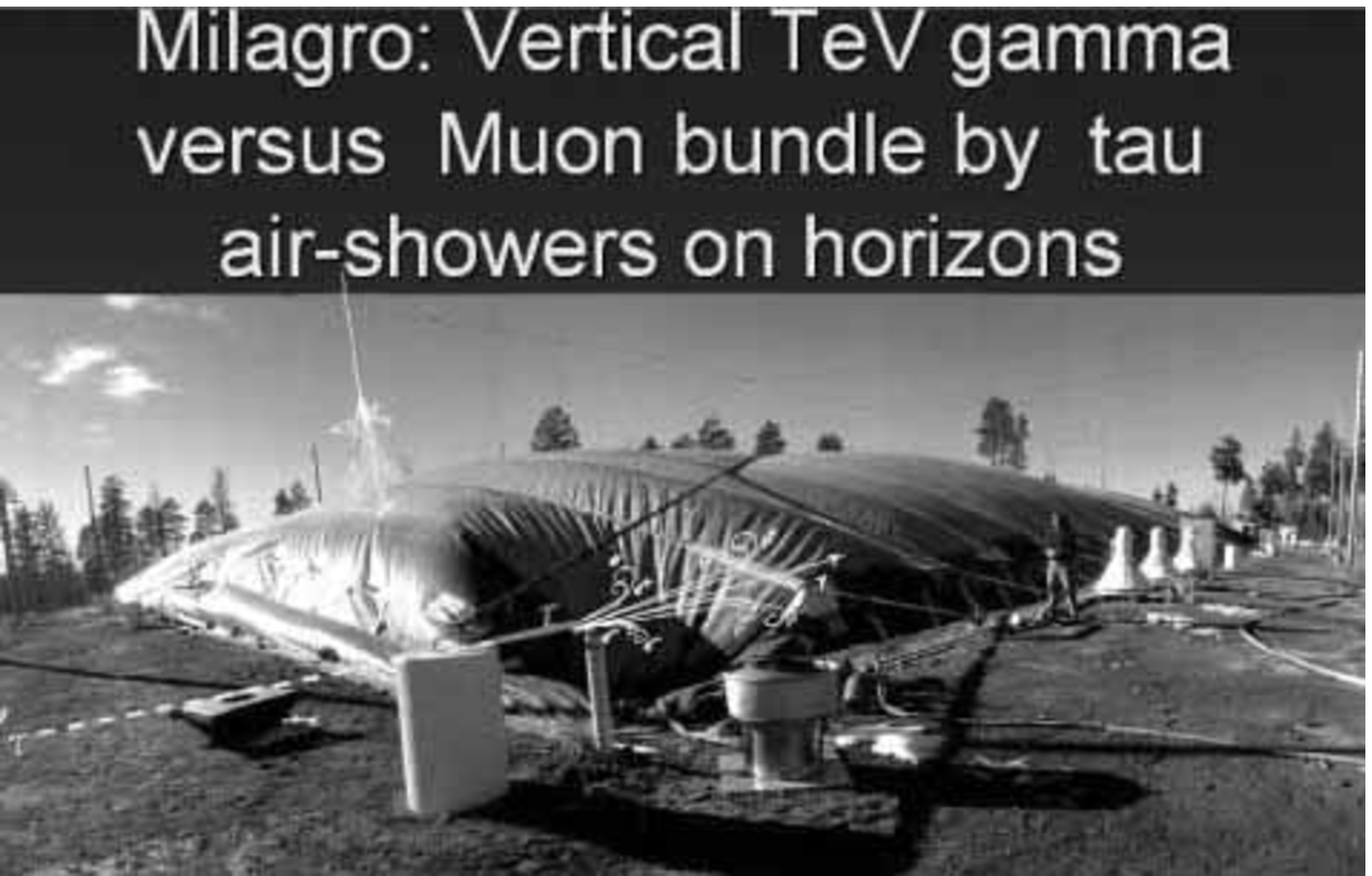}
\caption {The possible UHECR horizontal or up-going Tau
air-showering on  Milagro (as well in correlated mode, to nearby
Milagrito) while being  a TeV gamma detectors: GRB or an active BL
Lac at horizons, making nearly $1-3\%$ of GRB,SGRs,BL Lac events ,
might play a role in shining and tracing muon bundles in the
Milagro pool waters. As a first estimate, assuming an effective
area of few $10^3 \cdot m^2$ we foresee one or a few events of
Upward muon bundles associated to Tau Air-Showers each year,
depending on the trigger, the  threshold and geometry.
}\label{fig:fig1}
\end{figure}
%%%%%%%%%%%%%%%%%%%% fig 6 %%%%%%%%%%%%%%%%%%%%%%%%%
  \subsection{Milagro}
   The existence of huge pools at peak mountains as Milagro and
   smaller Milagrito, offer an exiting laboratory to verify
   (besides TeV gamma backgrounds): the muon horizontal fluxes at
   horizons; the muon bundle density, flux and structures (in
   comparison with sea-level Nemo-Decor data, see \cite{Decor},\cite{NEVOD}); finally there is
   the possibility to discover Up-going muon bundles, whose
   existence maybe indebt only to Earth Skimming (Uptaus)
   Air-showering.
 \subsection{AUGER}
 As in figures and in captions the Auger experiment offer a unique
  occasion to Horizontal Tau possibly from the West side toward
  the AUGER detector; to optimize the ability to disentangle these
  events one should first observe the Ande shadows (at $87-90^o$),
  zenith angles by a simple asymmetry East-West UHECR showering,
  see\cite{Fargion1999},\cite{Fargion 2002a}\cite{Fargion2004},\cite{Miele et. all05}.
  Within the first year of full operation the shadow $must$ be
  seen. Later on, within the same solid angle of $\simeq 2\cdot 100 = 6\cdot 10^{-2} sr.$
 two event a year by tau Air-showers (via GZK neutrino flux)
 $might$ be  very probably observed see \cite{Fargion2004}.

%%%%%%%%%%%%%%%%%%%% fig 8 %%%%%%%%%%%%%%%%%%%%%%%%%
\begin{figure}
\centering
\includegraphics[width=.6\textwidth]{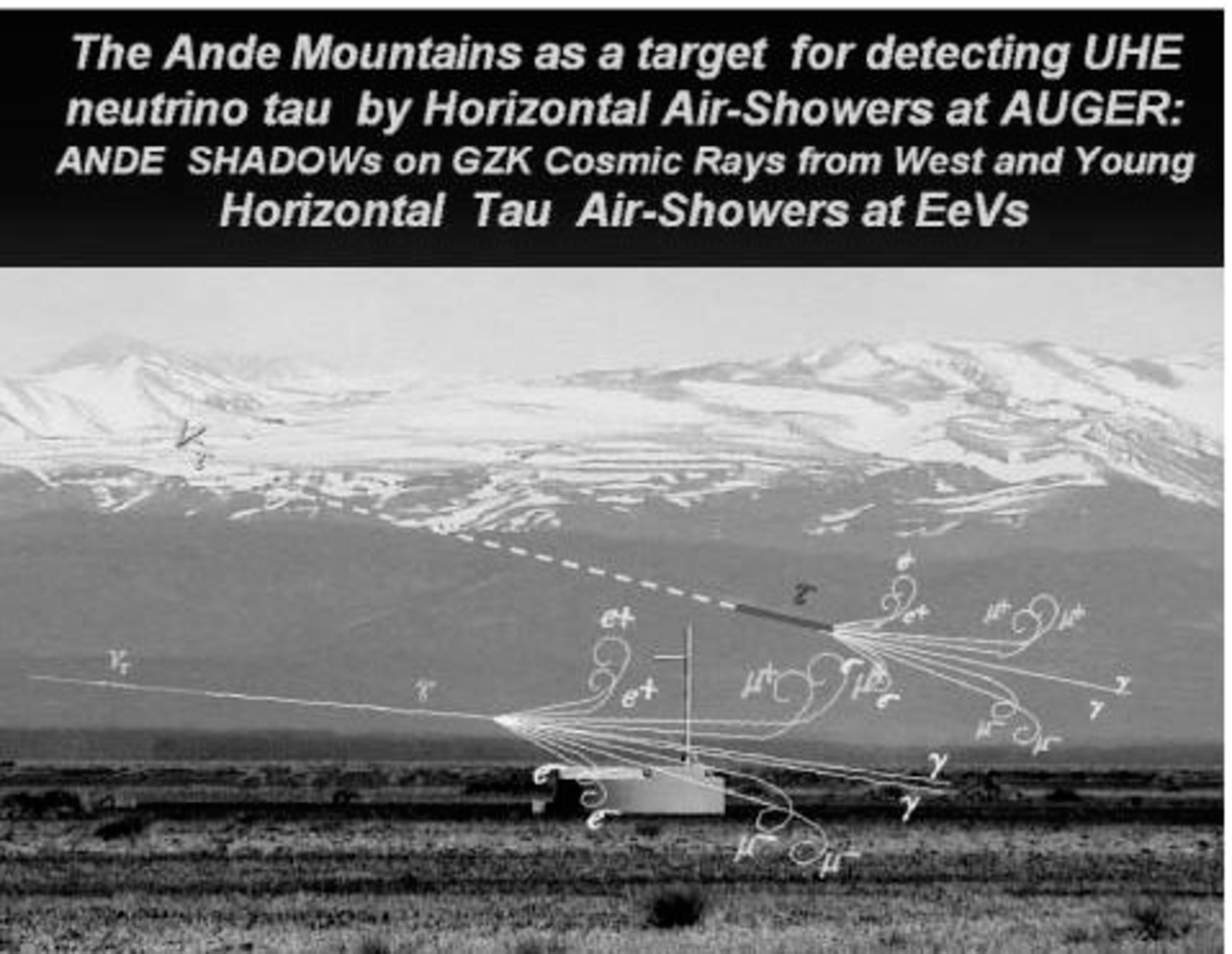}
\caption {The long Ande chain mountain is offering a unique wide
screening shadows (for UHECR) (opening angle $2-3^o$) and an ideal
beam dump (for PeVs-EeVs tau neutrinos) to AUGER array detector.
Inside this shadows, that may be soon  manifest, rare (a few a
year), but quite guaranteed horizontal tau (by GZK neutrino
fluxes) air-shower that might be open Neutrino Astronomy
windows.}\label{fig:fig1}

\includegraphics[width=.6\textwidth]{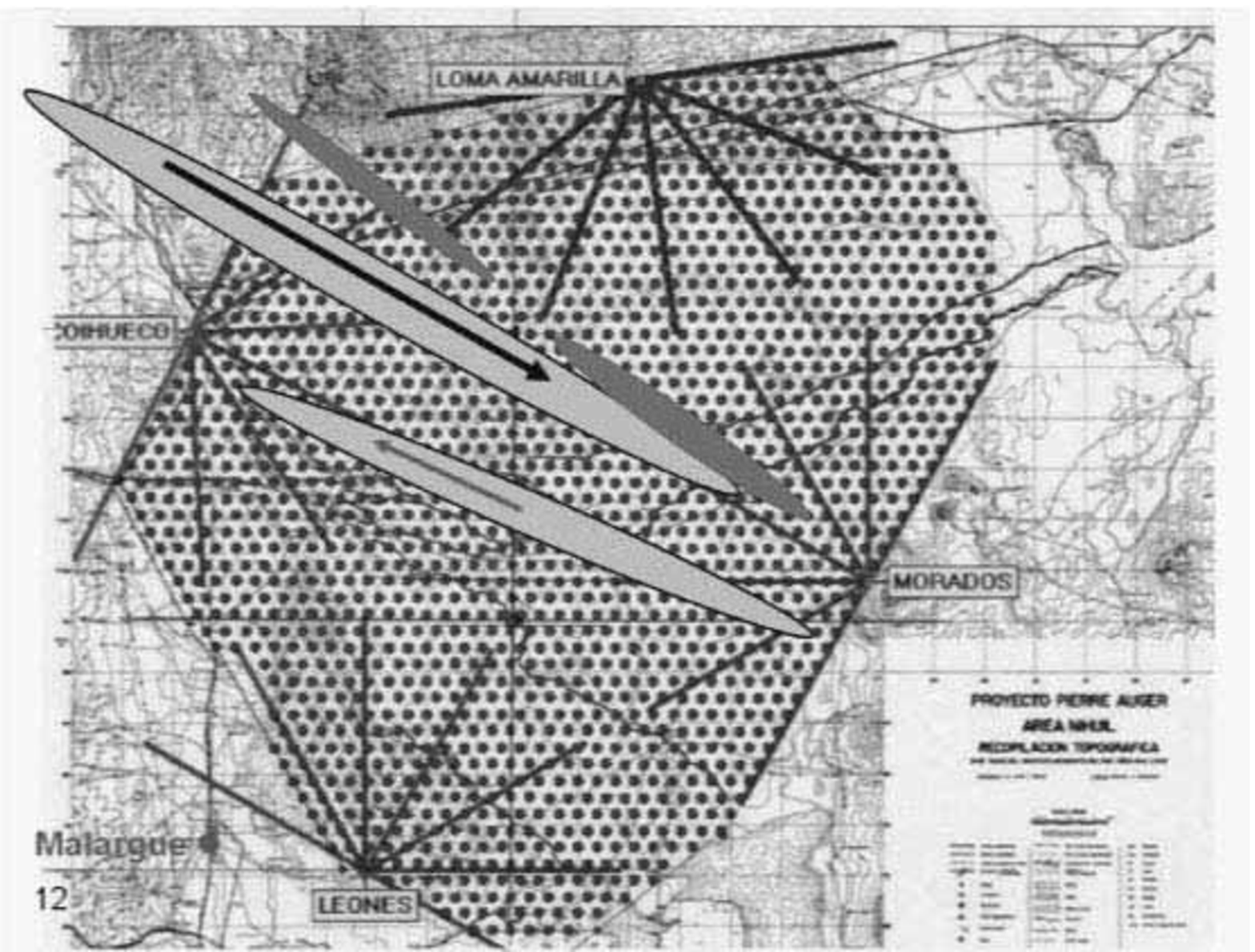}
\caption {Inclined Horizontal Air-Showers able to trigger $both$
Auger tanks $and$ Fluorescence telescopes , while being in the
same axis. This tecnique, as long the author knows, has never been
used to better disentangle horizontal Air-Showers. It may be an
ideal detector to observe Tau Air-Showers from the Ande. Their
events will populate the forbidden area of large zenith angle at
horizons. For this reason it will be useful to: a) enlarge the
angle of view of Coiuheco (as well as Loma Amarilla and Leones
station) toward the Ande;
 b) to eliminate any optical filter for Cherenkov lights in those directions ;
 c) to open a trigger between the Array-Telescope, or Telescope-Telescope in Cherenkov blazing mode; d) to try all $4$
telescope Fluorescence connection in Cherenkov common trigger-mode
 along all the $6\cdot 2 = 12$ common arrival directions.  Similar
connection along the $360^o$ view of stereoscopic HIRES
telescopes, might be already done testing along  their common
horizontal $2$ axis   air-showers (from PeVs up to EeV energy)
with  high rate (tens-hundreds events a night)
 and great angular accuracy. In the picture some possible inclined UHECR
 events shining both array detectors and  (by Cherenkov lights) Fluorescence Station;
  possible twin separated ovals arise by geomagnetic bending.}\label{fig:fig1}
\end{figure}
%%%%%%%%%%%%%%%%%%%% fig 6 %%%%%%%%%%%%%%%%%%%%%%%%%

    The AUGER Fluorescence detector may enlarge their view also
  toward the Ande, offering an ideal screen capturing Ande-Tau
  Showers in horizontal tracks at best. The possibility to use
  inclined air-shower Cherenkov lights hitting the Fluorescence
  detector maybe exploited. Multi-telescope coincident Cherenkov
  detection (while being nearly on axis) of horizontal air-showers
  maybe applied in all the $12$ common directions ($4\cdot3$) in AUGER (and $2$ for
  stereoscopic HIRES).
 \subsection{Space Station Crown Arrays }
  From the Space there is the most appealing location to search
  for Horizontal High Altitudes Showers and Horizontal or Vertical
  Tau-Air-showers (Hortau-Uptau). This project is still
  preliminary. The crown-array maybe $both$ detecting (tens, hundred keV) gamma secondaries
 (as well as rarer hundred GeV lepton pairs)
  as well as Cherenkov lights due to far Hadron and Gamma primary High
  Energy Cosmic Rays showering from Earth.  The array maybe at
  PeVs-EeVs energy equivalent to few-hundred km. mass neutrino
  detectors, depending on the telescope sizes and gamma array
  area.See \cite{Fargion2001},\cite{FarCrown}.
\subsection{EUSO }
The project of a telescope facing down-ward the Earth and catching
the UHECR has been delayed to the end of the century. However the
idea may offer a way to discover beamed horizontal HorTaus at tens
EeV energy showering at high altitudes. Few events , $4-6$, might
be observed each year. The EUSO mass equivalent due to
Earth-Skimming  \cite{Feng2002}, or HorTaus \cite{Fargion 2002a} neutrinos is nearly $100 sr km^3$ \cite{Fargion2004} water equivalent,
even taking into account the $10\%$ duty cycle of the EUSO
activity.See \cite{Fargion2004}.

%%%%%%%%%%%%%%%%%%%% fig 8 %%%%%%%%%%%%%%%%%%%%%%%%%
\begin{figure}
\centering
\includegraphics[width=.5\textwidth]{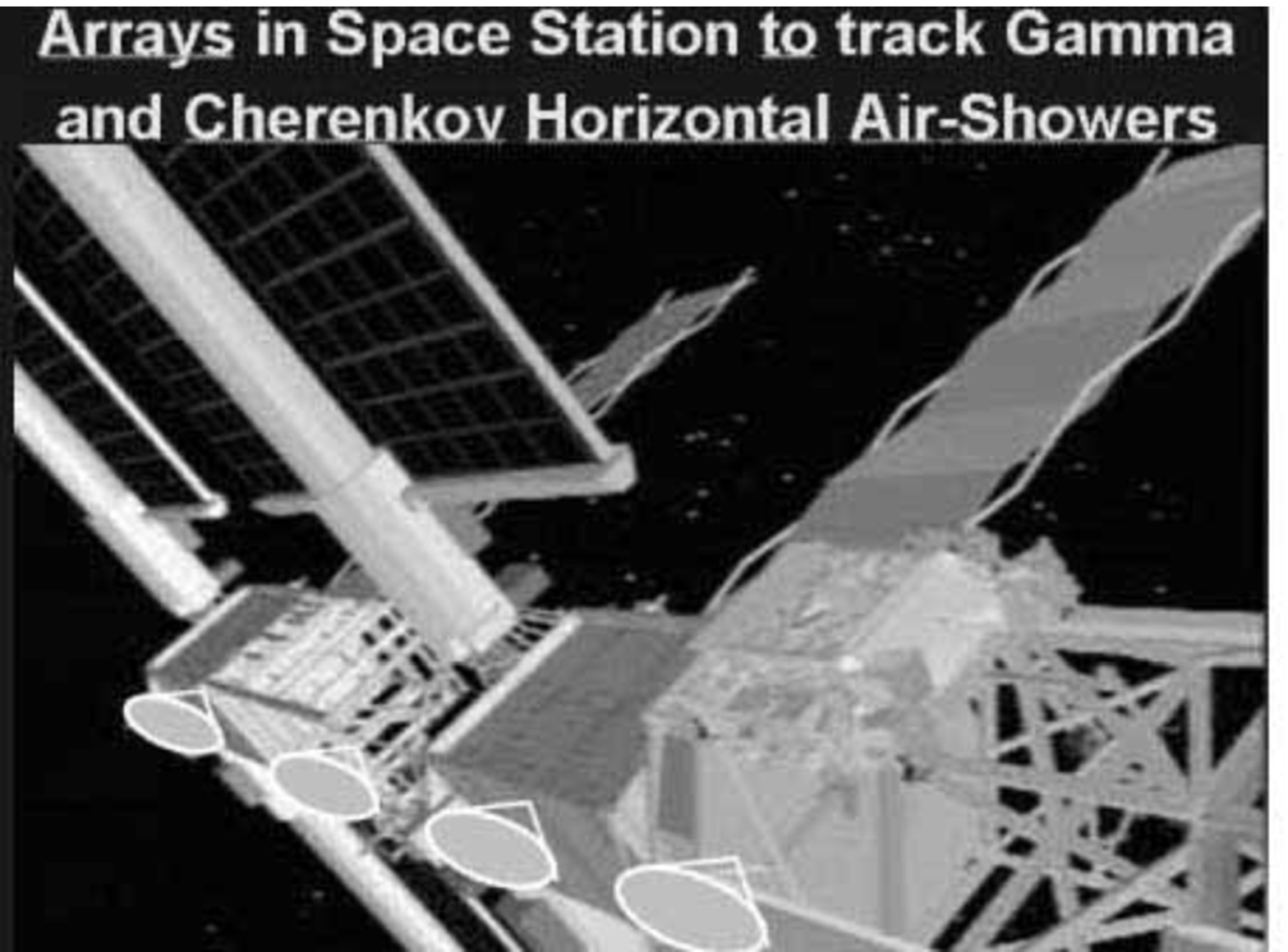}
\caption {Space Station constructed and armed with a Telescope
Cherenkov Array and a Gamma Array spread array able to disentangle
gamma flashes and arrival lights, from the Earth edges. The
possible Tau neutrino nature is imprinted by the arrival direction
$below $ the Earth horizons, while the UHECR showers arise at the
high Atmosphere (Albedo) edges above. The duration of the signal
(micro-second to millisecond), as for Terrestrial Gamma Flashes,
is the signature of these Up-going Air-Showers, steady ones are
the signature of Gamma TeVs-PeVs air-showering sources. The
threshold depends on the Telescope and Gamma detector areas; even
the distances from Space Station are nearly $100-200$ larger than
vertical TeV-PeV air-showers on Earth, the beaming is $14-20$
smaller, with negligible absorption, making a $2$ meter square
Cherenkov telescope able  in principle to observe TeVs gamma
sources. )}See \cite{FarCrown}\label{fig:fig1}

\includegraphics[width=.6\textwidth]{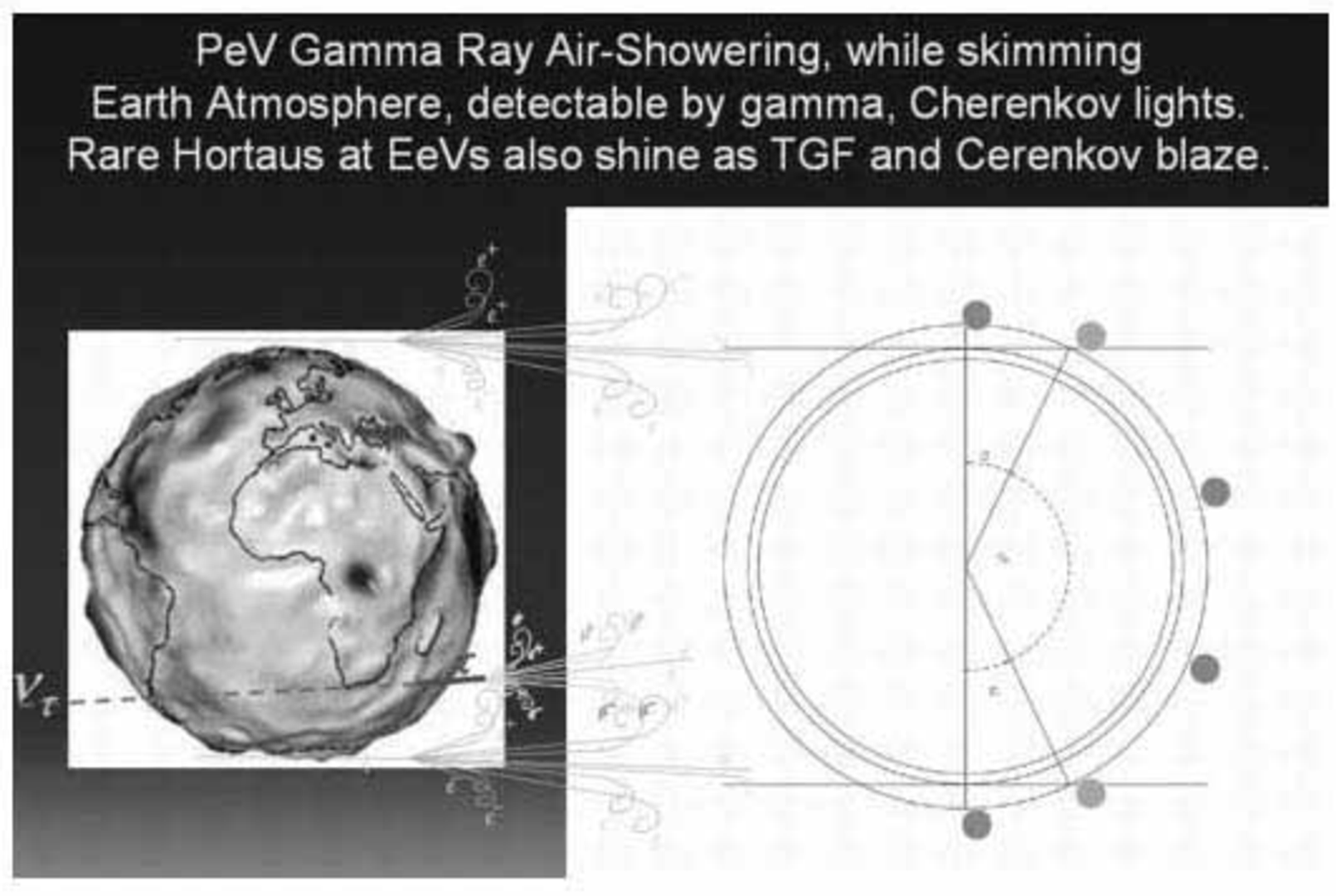}
\caption {Gamma and UHECR Air-Showering (High Altitude Air
Showers, HIAS), in Space versus Tau Horizontal and Vertical
(HorTau-UpTau) air-showers: while the Atmosphere Earth Skimming at
TeVs-PeV gamma photons are showering in high altitude atmosphere
($35-40$ km.), at extreme low air density (about or below $1\%$ os
sea level), their Cherenkov beam is about $0.1^o$ at $36$ km
altitudes. Their air-showering  may blaze from $2000$ km far away
 leading to brief, $persistent$  (about two in gamma-X rays or tens second in optical duration),
  transient gamma and Cherenkov  point-like flares, detectable by the present and future Gamma Satellites
  array. Their exact arrival angle maybe  monitored within narrow angles by a Crown
   Cherenkov Telescopes on the Space Station toward the terrestrial edges.
    Similar less rare $gamma-air-skimming$ from Tens GeV to hundred GeVs events,  might indeed be
   already hidden in some old records of  BATSE catalog, labelled  as $electrons$ or $particle$ events.
   Indeed their puzzling signature of the satellite rise and dawn orbit, as in figure, are notable.
     The well known discover of  brief up-going Terrestrial Gamma Flash
   by BATSE ($1991-2000$) maybe, on the contrary be indeed indebt to  Earth-Skimming EeV
    neutrino showering (Hortaus) or PeVs Up vertical showers (Up-taus), or to UHECR air-skimming the
    Earth atmosphere}see \cite{FarCrown}.\label{fig:fig1}
\end{figure}
%%%%%%%%%%%%%%%%%%%% fig 6 %%%%%%%%%%%%%%%%%%%%%%%%%

%%%%%%%%%%%%%%%%%%%% fig 8 %%%%%%%%%%%%%%%%%%%%%%%%%
\begin{figure}
\centering
\includegraphics[width=.6\textwidth]{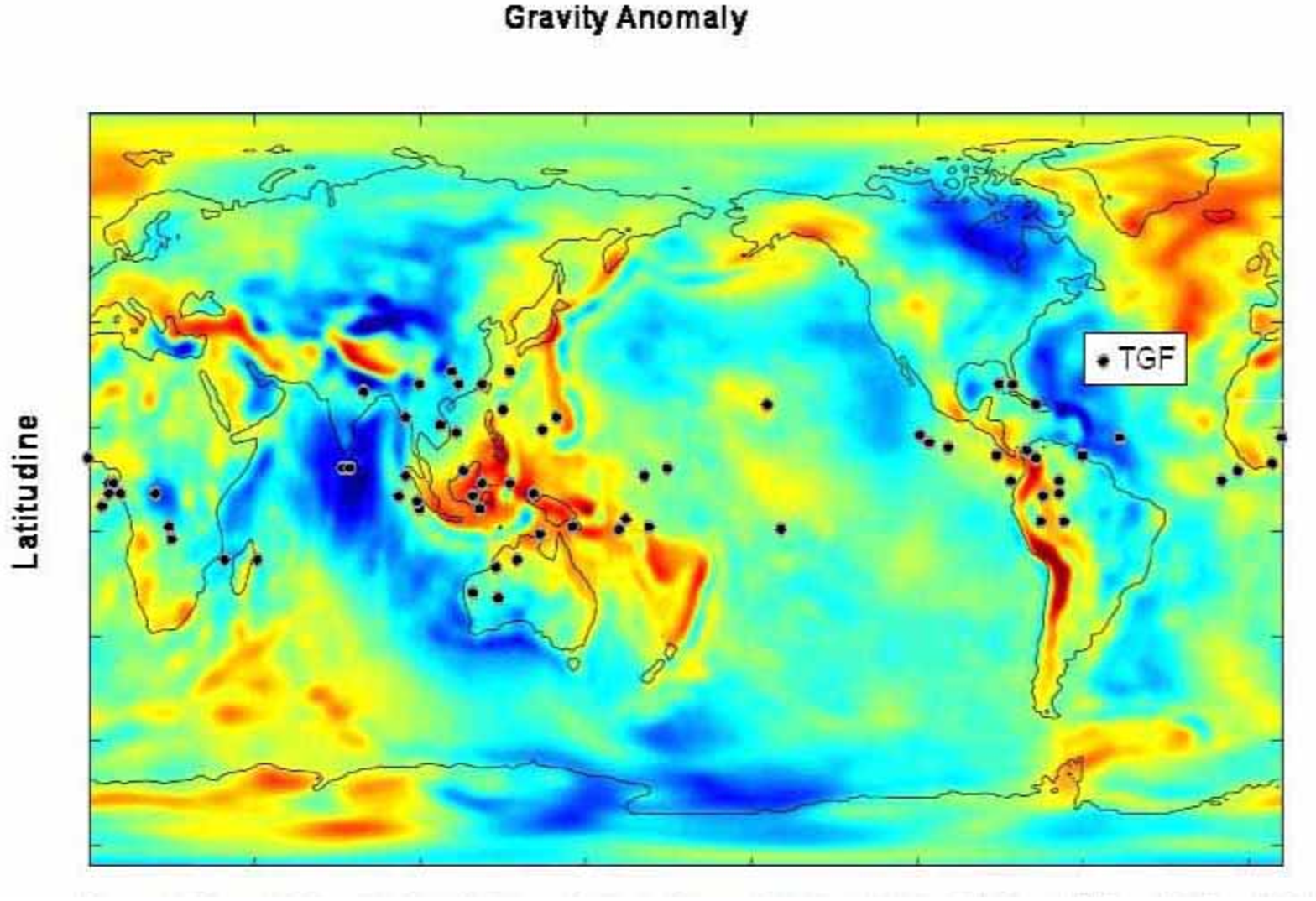}
 \caption {The apparent correlations between Earth
crust contrast, gravity anomaly and the observed location of
Terrestrial Gamma Flashes observed by BATSE in last decade (and by
RHESSI last two years). The overlap of the TGF events with maximal
terrestrial mass density contrast (Mountain chains, sea-islands )
(in the equatorial belt where BATSE-Compton trajectory laid),
favors a common origin of TGF and tau-airshowers. }See  \cite{Fargion 2002a},\cite{Fargion2004}.\label{fig:fig1}

\includegraphics[width=.6\textwidth]{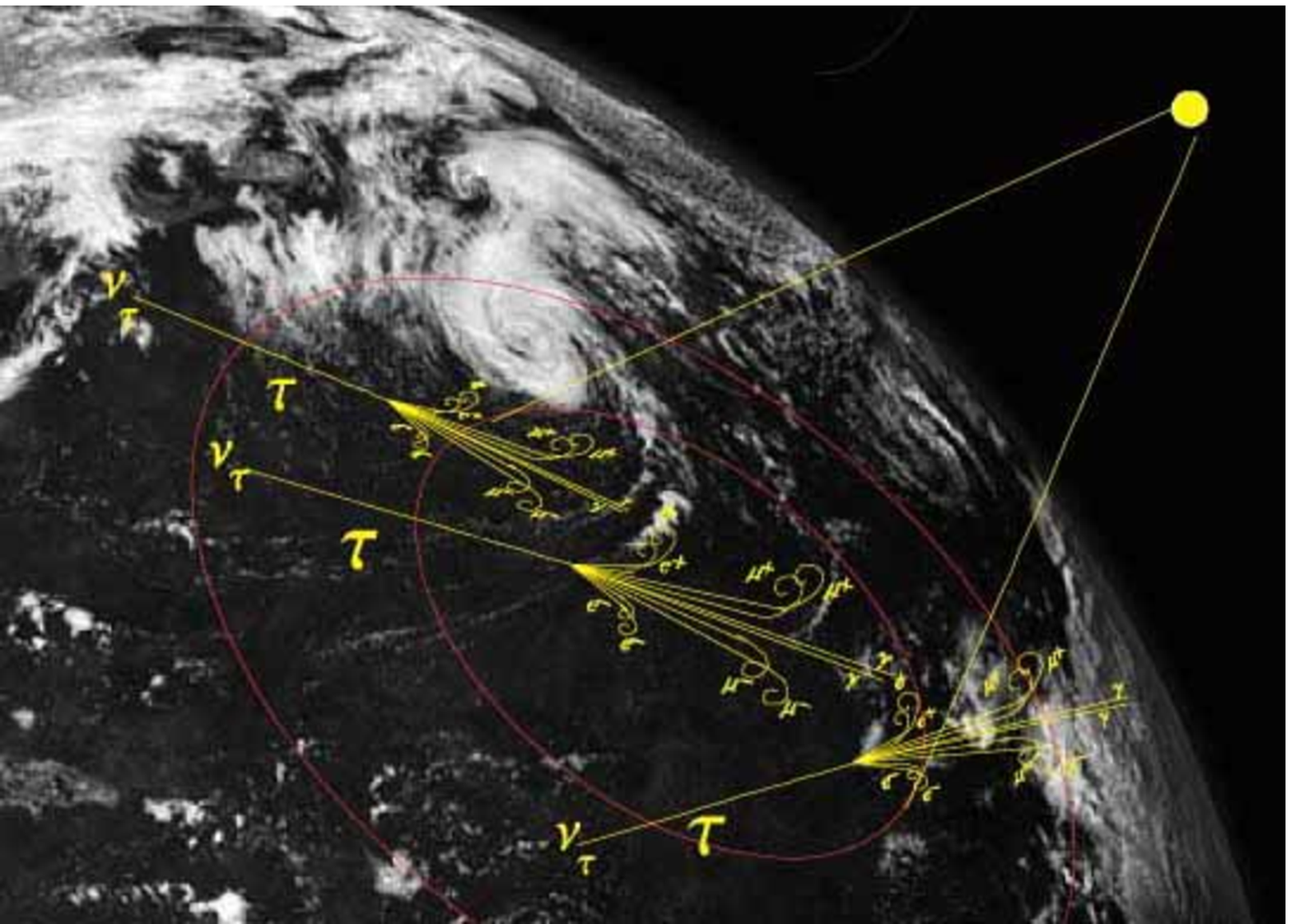}
 \caption {The up-going horizontal air-showers whose longest (hundreds of km.) air-showers
 might be detectable by future Euso project; the project would reveal thousands oh UHECR
 mostly downward events, as well as hundred of horizontal C.R.
 airs-showers, whose beam angle is extremely small, because low air density.
 Within these down-going UHECR air-showers there are 4-6 event a year originated
 within a wider field of view. One of his greatest proponent and great
 scientist,that with Prof. Linsley discovered   UHECR  at GZK  edges,
 Prof. M.Livio, sadly has very recently missed.}See  \cite{Fargion2004}\label{fig:fig1}
\end{figure}
%%%%%%%%%%%%%%%%%%%% fig 6 %%%%%%%%%%%%%%%%%%%%%%%%%

\subsection{BATSE}
Old generation of Gamma satellite in orbit last decade made (with deep discovers by Beppo Sax) most
of our view in gamma astronomy. Present and next generation (Swift,Glast) will enlarge the
EGRET astronomy by deeper views. The same skimming C.R. or Albedo and Air-Shower
tracing by UHECR (PeVs-EeVs) will naturally arise \cite{Fargion2001}.
\subsection{ASHRA }
Three  Fluorescence detectors, in a similar way as
  AUGER telescopes, are monitoring from the top mountains of the Hawaii island the
  inner area; their detection maybe  greatly enhanced by tracing and calibrating higher altitude HIAS and
  facing the Earth edges, searching the HorTaus at ocean Horizons.

%%%%%%%%%%%%%%%%%%% fig 8 %%%%%%%%%%%%%%%%%%%%%%%%%
\begin{figure}
\centering
\includegraphics[width=.4\textwidth]{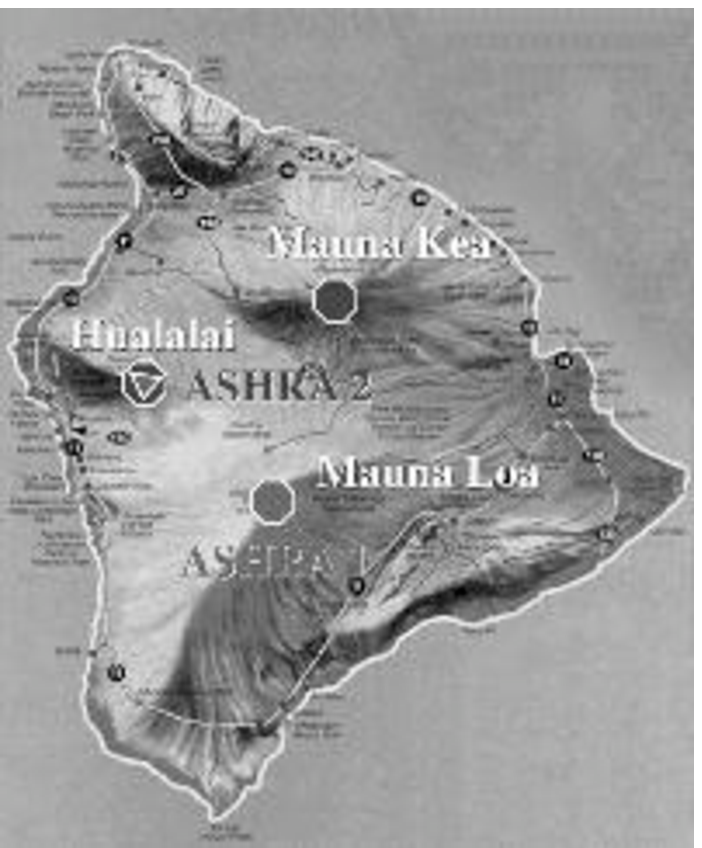}
 \caption {The present ASHRA experiment in
Hawaii leading the rush of Horizontal Tau Air-Showers by
Fluorescence Telescope Array on the top mountains.}

\includegraphics[width=.5\textwidth]{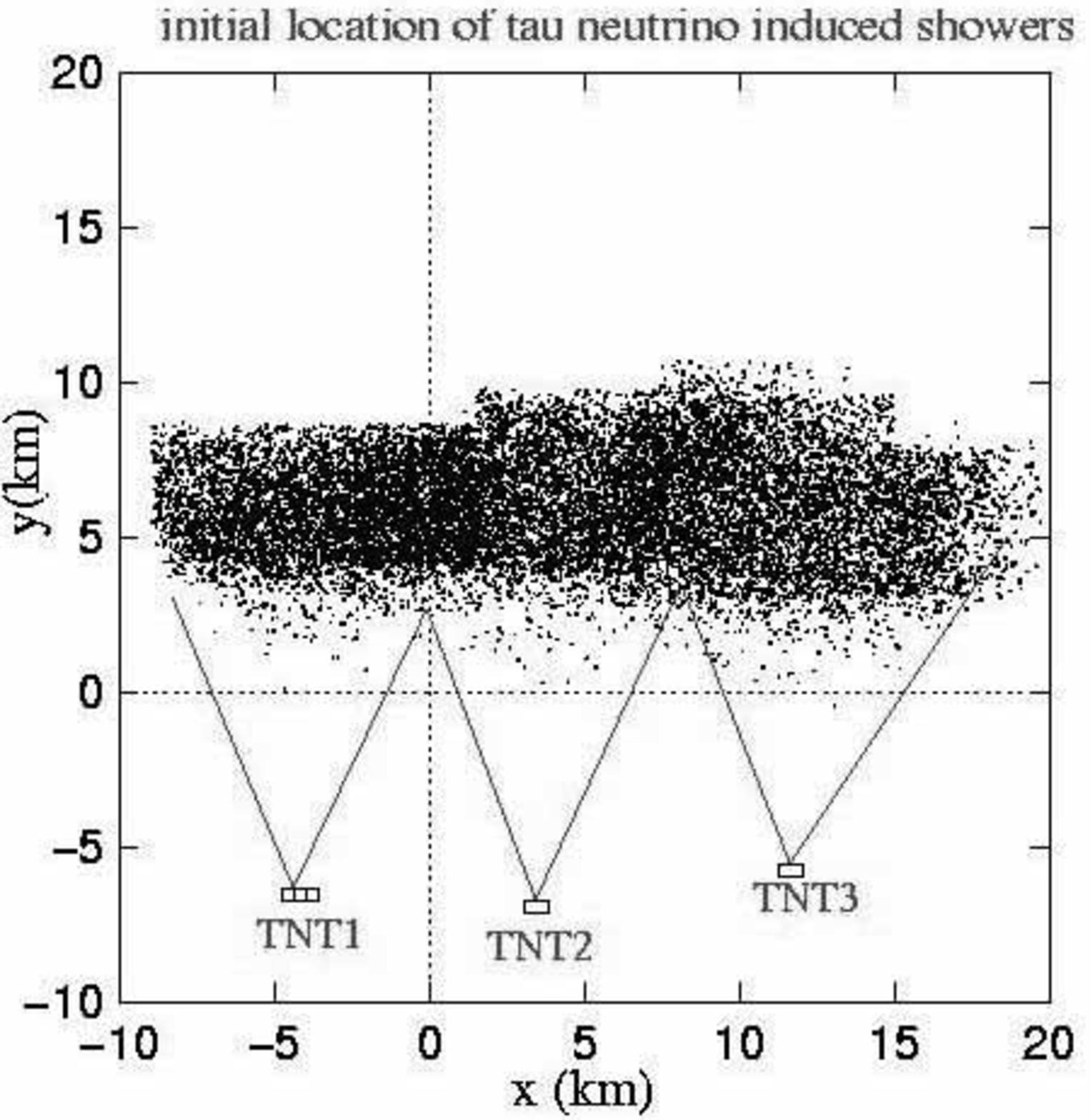}
 \caption {The project experiment in
UTAH nearby (ten km.) a   cliff searching for Cosmic Ray Tau
Neutrino, CRTNT, by Fluorescence Telescope Array.}\cite{Cao}

\includegraphics[width=.5\textwidth]{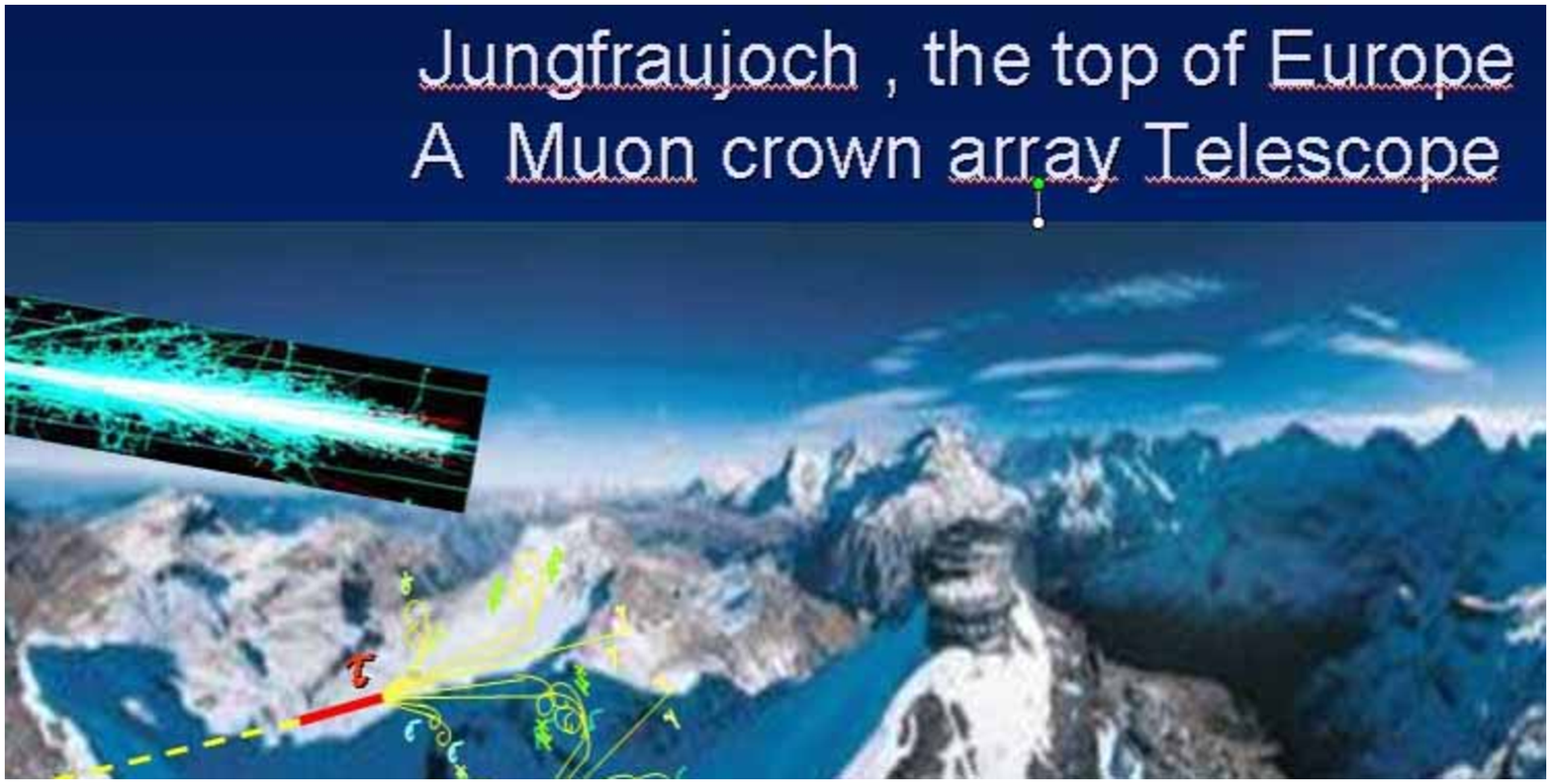}
 \caption {A novel project experiment in
Swiss top-mountain   for Horizontal Cosmic Ray Tau
Neutrino as well as horizontal muon fluxes
.}See \cite{Iori04}; \cite{Grieder01}.
\end{figure}
%%%%%%%%%%%%%%%%%%%% fig 6 %%%%%%%%%%%%%%%%%%%%%%%%%

\subsection{Jungfraujoch} The existence in the top Europe turistic station
of scientific facilities and fast transport made possible a first
test of  Muon Telescope Array prototype at horizons site. The
proposal of a larger area and more numerous detector is in
progress and it may compete with Cherenkov telescope also because
of light noise independence of the scintillator array.
\subsection{CRNT}
The proposal of a Fluorescence array within the cliff shadows is
going to be considered in Utah and-or in China. The PeV detection
will be possible by low noise light location and Cherenkov  aided
discovering techniques.
 \subsection{Shalon}
 A Russian proposal leaded by Cherenkov TeV-Telescope is already
 looking  from  the mountains terrestrial
 targets in search of eventual Tau air-showers, finding already a statistics
 on Albedo air-Showers and relevant first calibrations.
\subsection{Eolic Array}
 On the top of the mountains small crown  arrays of detectors may be
 mounted on the eolian energy stands. The usual sites ( mountain)
 the power supply, the two different hight on the same element may
 offer a useful place to build an horizontal Air-shower detector,
 in wild areas and spread surfaces \cite{FarCrown}.

\section{The Veritas and  Magic views  of Tau Air-Showers at horizons}

Cherenkov gamma Telescopes as last Veritas and  MAGIC ones at the
top of a mountains are searching for tens GeV $\gamma$ astronomy.
The same telescope at zero cost in cloudy nights, may turn (for an
bending angle $\simeq 10^{\circ} $) toward terrestrial horizontal
edges, testing both common PeVs cosmic ray air showers, muon
secondary noises and bundles as well as upgoing tau air-showers.
Indeed the possible detection of a far air shower is enriched by:
\begin{enumerate}
  \item early Cerenkov flash even dimmed by atmosphere screen
  \item single and multiple muon bundle  shining   Cerenkov rings or
  arcs inside the disk  in time correlation
  \item muon decaying into electromagnetic in flight making
  mini shower mostly outside the disk leading, to lateral correlated gamma tails.

\end{enumerate}
  We estimated the rate for such PeVs-EeV events each night,
  finding hundreds event of noises muons and tens of bundle
  correlated signals each night \cite{Fargion2005}. Among them  up-going Tau
  Air-Showers may occur very rarely, but their discover is at hand for
  dedicated $360^o$ crown Arrays \cite{FarCrown}(and arrays of these crowns) in correlation among
  themselves and scintillator detectors.

%%%%%%%%%%%%%%%%%%%% fig 8 %%%%%%%%%%%%%%%%%%%%%%%%%
\begin{figure}
\centering
\includegraphics[width=.7\textwidth]{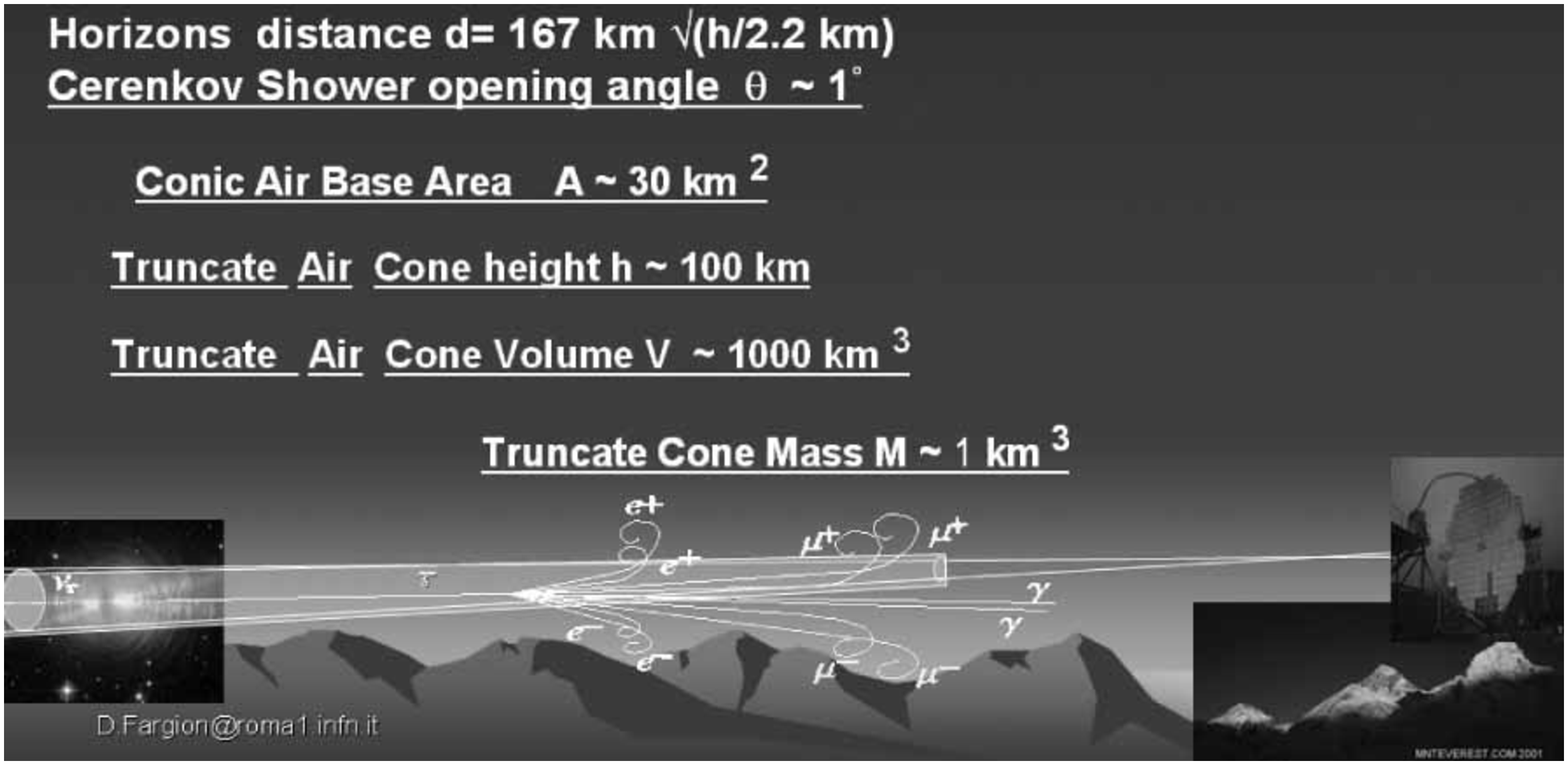}
\caption {The possible horizontal air-showering by a GRB or an
active BL Lac , whose UHE anti-electron neutrino might resonance
with air electrons at Glashow PeVs energies (or in
 Tau air-showers at higher energy), making  nearly $3\%$ of these GRB,SGRs,BL Lac
 sources  laying at horizons for Magic  Telescopes. The mass observed , as estimated in figure, within the
 air-cone exceed the $km^3$ water mass, even if within a narrow solid angle ($\simeq 4\cdot 10^{-3}$ sr.) }See \cite{Fargion2005}\label{fig:fig1}

\includegraphics[width=.7\textwidth]{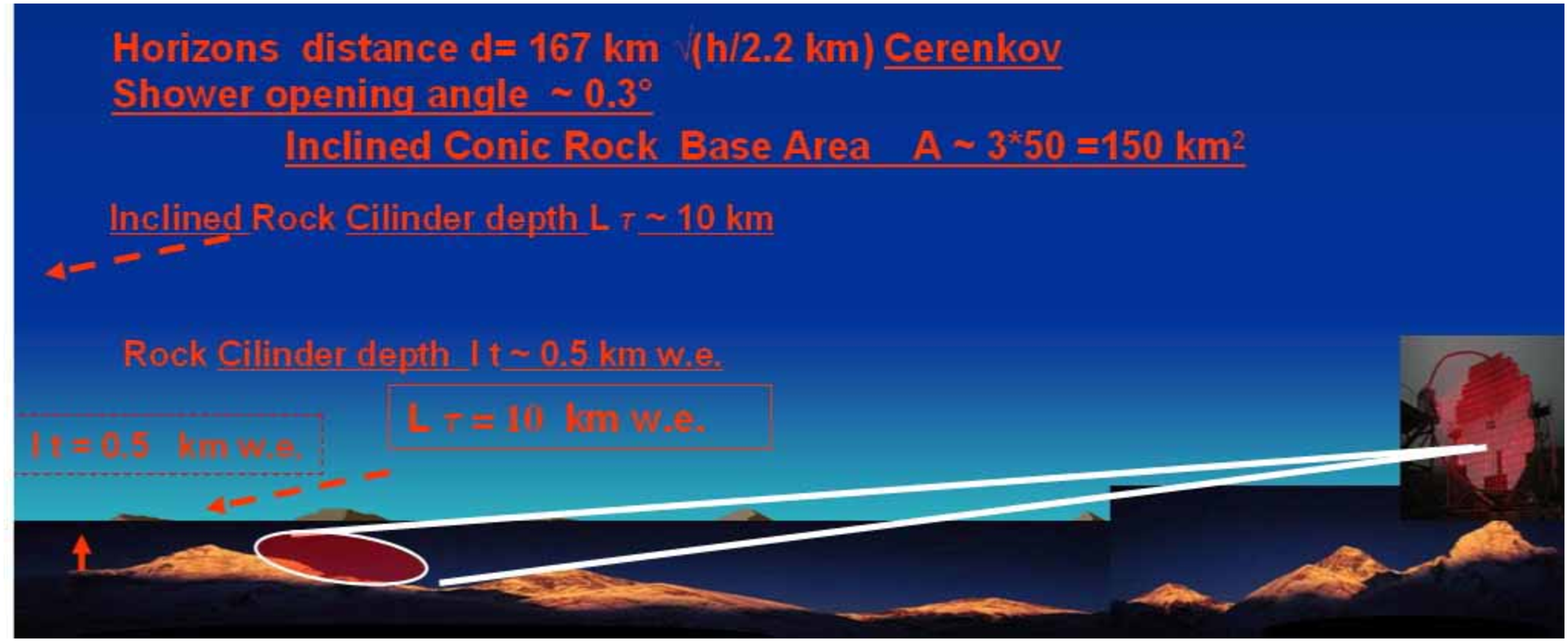}
\caption {As above  EeVs tau  are originated in the Earth crust
 and while escaping the soil are testing $\sim 70-100 km^3$ volumes; later
 UHE tau may decay in flight and may air-shower loudly toward Magic telescope, within an area of
 few or tens $km^2$.}See \cite{Fargion2005}.\label{fig:fig1}
\includegraphics[width=.7\textwidth]{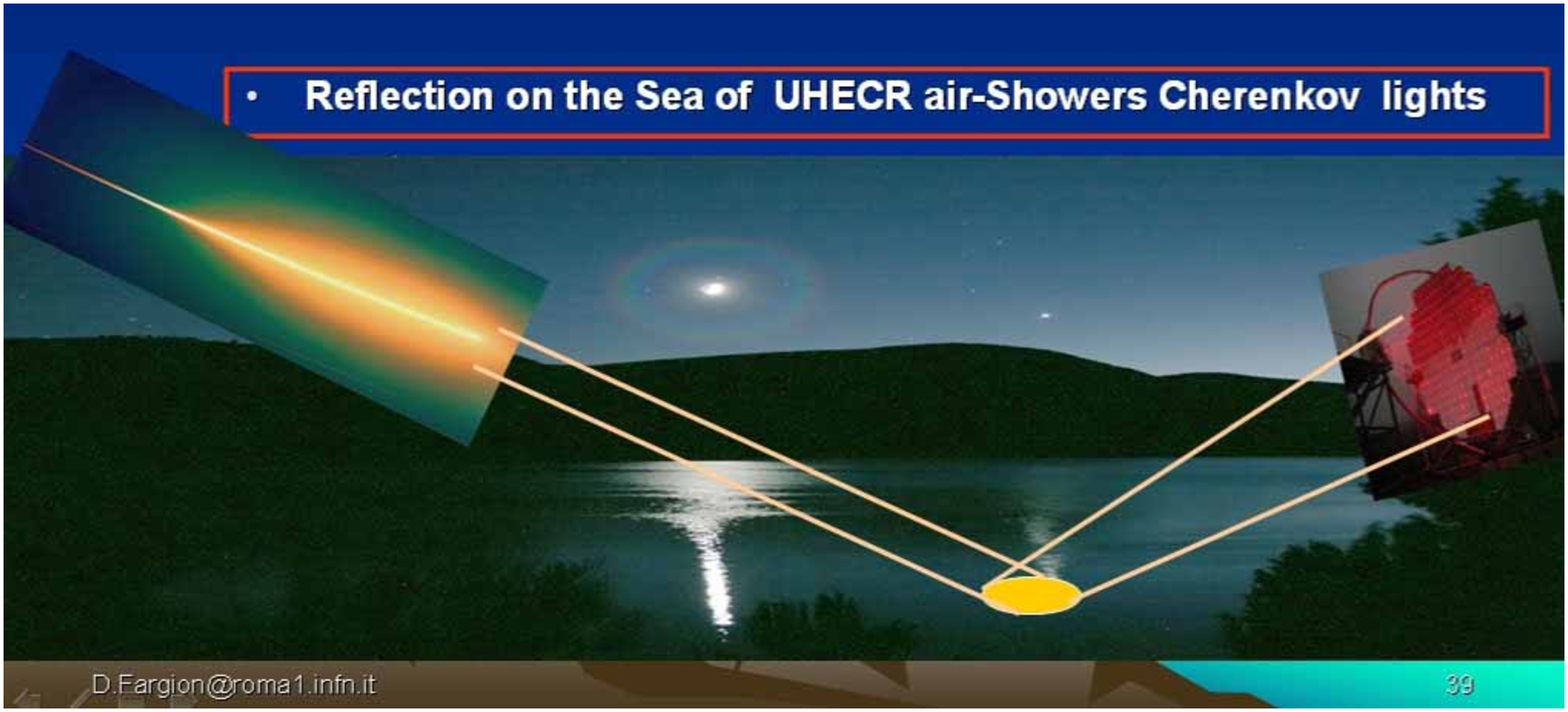}
\caption {The possible inclined UHECR air-showering on Magic
facing the sea side. Their detection rate is large (at zenith
angle $80-85^o$) (tens or more a night)
 nearly comparable with those  at zenith angle  $87^o$ already estimated;
  these mirror UHECR shower , widely spread in oval
  images on the sea (depending on the sea wave surfaces), their presence
is an useful test for Magic discovering of point source PeV-EeV
UHECR air-showers at horizons.
 While previous configuration above horizons may correlate direct muon bundle and
Cherenkov flashes, these mirror events are polarized lights mostly
muon-free, diffused in large areas and dispersed in longer time
scales, mostly in twin (real-mirror-tail) spots. On the contrary
Up-going Tau air-showers from the sea are very beamed and thin and
un-polarized and brief.}See \cite{Fargion2005}.
\end{figure}
%%%%%%%%%%%%%%%%%%%% fig 6 %%%%%%%%%%%%%%%%%%%%%%%%%

%%%%%%%%%%%%%%%%%%%% fig 8 %%%%%%%%%%%%%%%%%%%%%%%%%
\begin{figure}
\centering
\includegraphics[width=.6\textwidth]{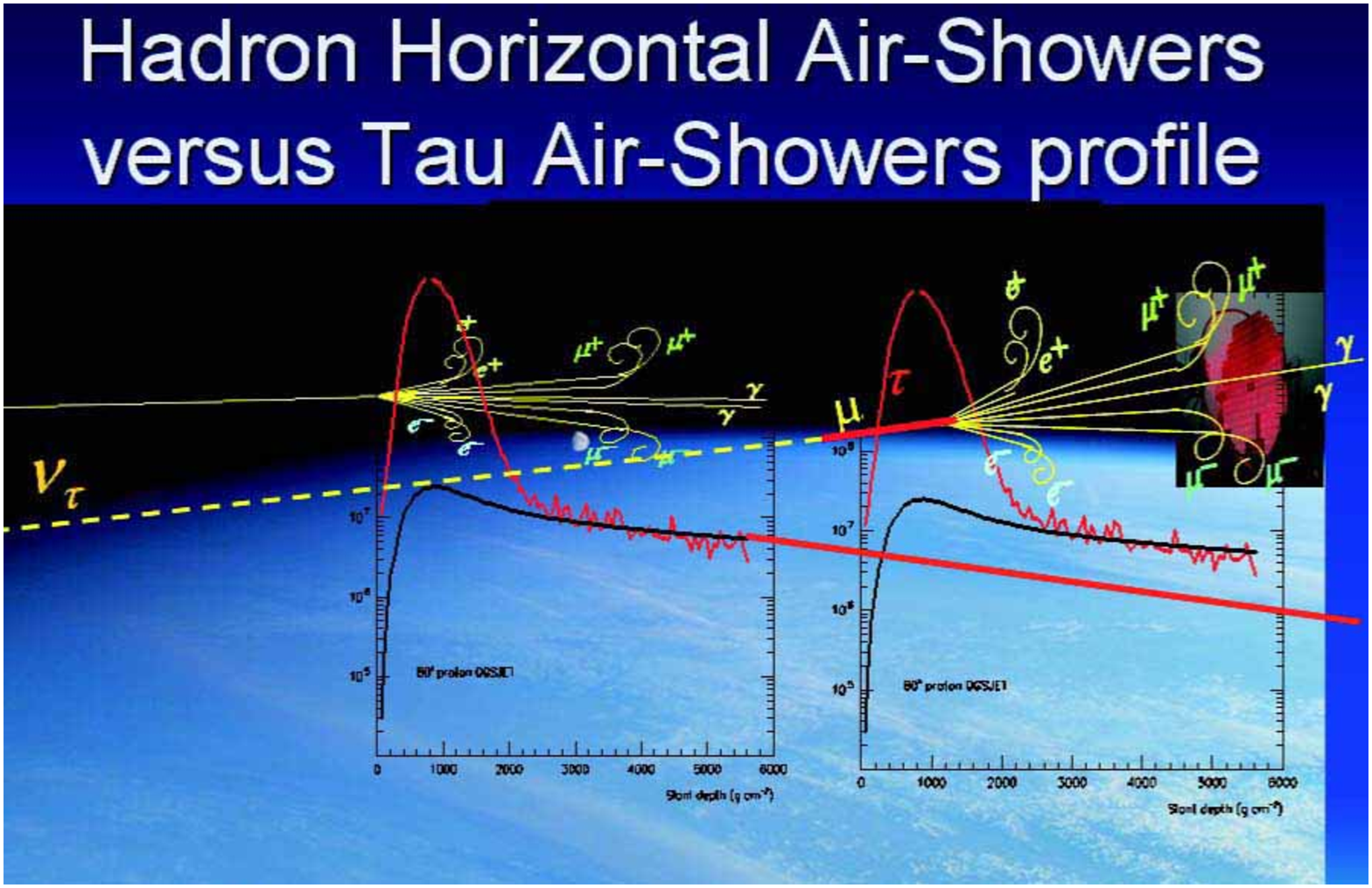}
 \caption {The  horizontal air-showers by far hadron differ to an up-tau air-showers,
 whose younger electromagnetic and muonic density is greater and much larger ;
  in the figure the two different
 signature of the flux densities assuming a Magic telescope observer (not in scale),
  and an ideal downward far nucleon and a nearby Tau EeV air-showers event.}\cite{Cillis2001}

\includegraphics[width=0.5\textwidth]{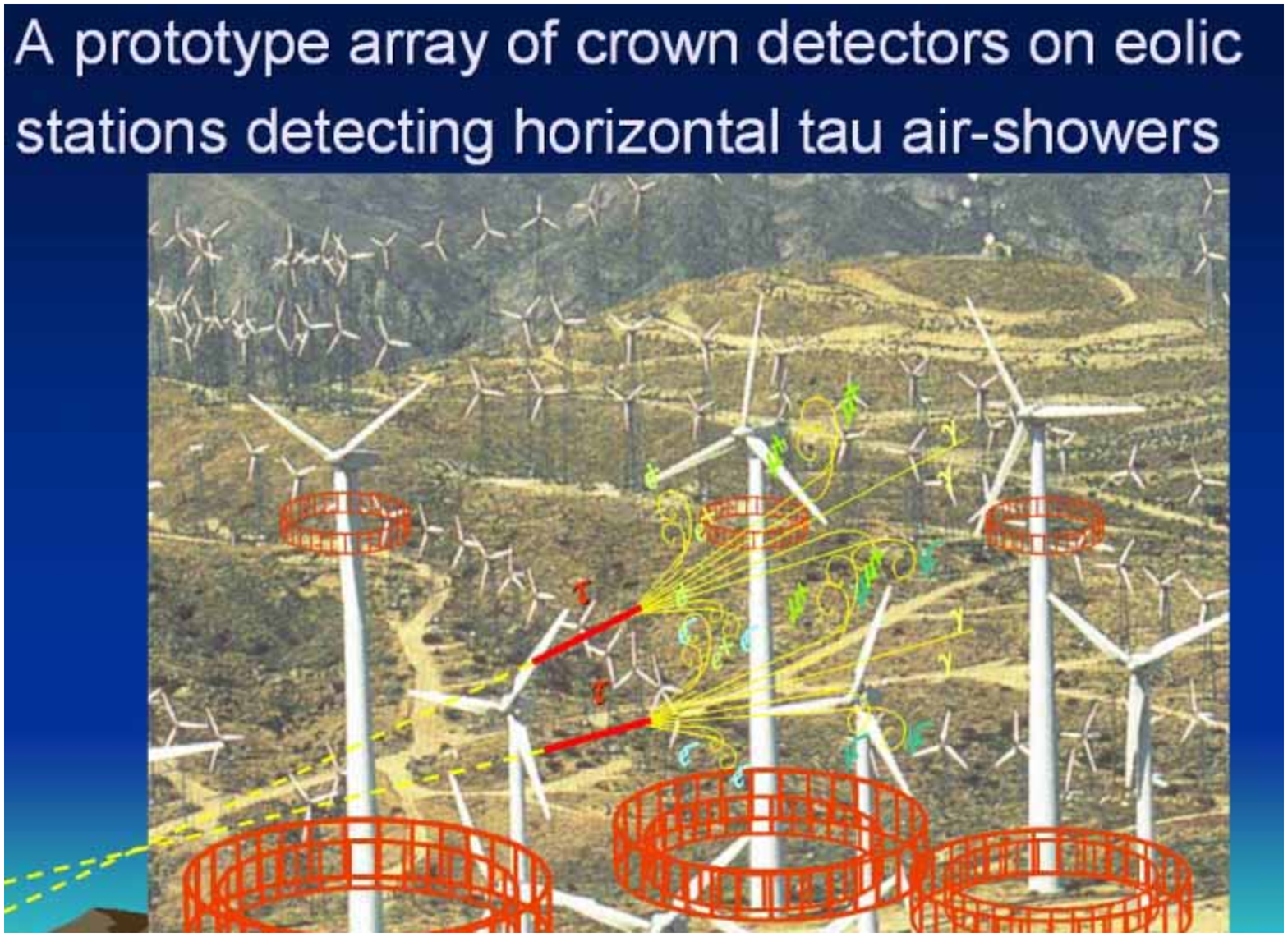}
 \caption {Ideal arrays of crown scintillators on wind eolic stations.}\label{fig:fig1}

\includegraphics[width=0.8\textwidth]{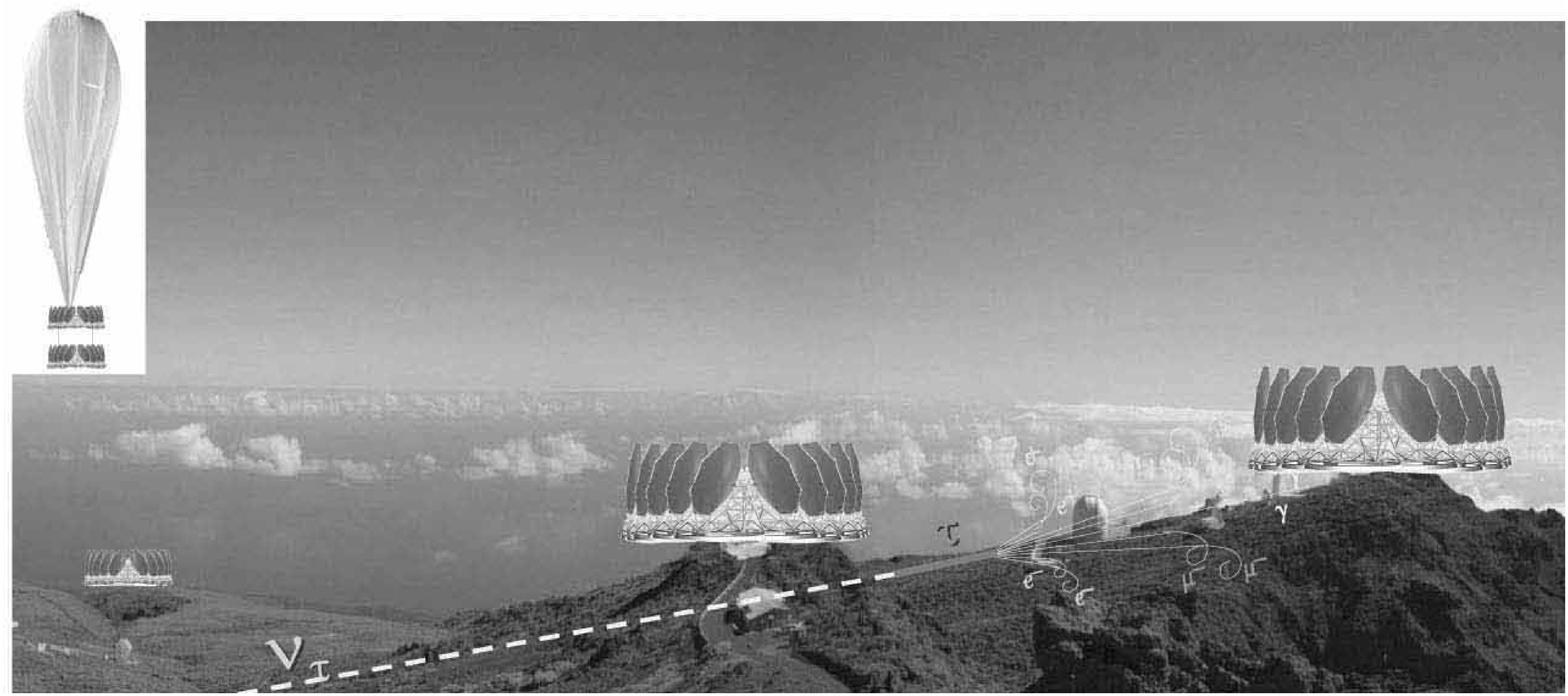}
 \caption {Ideal arrays of Cherenkov Crowns Telescopes in
 Canaries and an equivalent twin Crown Array Balloon  in flight; similar arrays maybe located
 in planes or satellites.}\label{fig:fig1}
\end{figure}
%%%%%%%%%%%%%%%%%%%% fig 6 %%%%%%%%%%%%%%%%%%%%%%%%%
The timing of these signals,their expected event rate at PeVs in a
night time of Magic at horizon ($87^o$ zenith angle ) and their
easy signature has been reported recently by Fargion
\cite{Fargion2005}.More over the very simple exercise of the estimate of
the air cone volume observed at the horizons by MAGIC shows a
value larger than $10^3Km^3$ corresponding to a mass volume larger
than $1km^3$ water equivalent.This volume at Glashow resonance
energies make MAGIC the must wide $\nu$ detector.The some estimate
even at smaller solid angle,below the horizon leads a large mass
for EeV neutrinos,encompassing volumes and masses as large as
$10^2km^3$.These detectors are active only within a narrow
view,but during peculiar rise and down of BL Lac,AGN or Crab like
sources, or in coincidence with GRBs along the horizon,the masses
enlisted are huge and relevant.To make the detection permanent and
in wider angle view the ideal crown array of MAGIC-like telescope
on circle and their twin or multiple array structure at few km
distance,will guarantee a huge capability to observe an event or
few event  of $\tau$ upgoing shower during a month,within a few
tens of UHECR above the edges.

\section{Conclusions}
Because muon tracks are  mostly  of downward atmospheric nature
the underground neutrino telescope are tracing rarer upward ones
mostly of  atmospheric origin;  higher energy upgoing  astrophysical
neutrino signals are partially suppressed by Earth opacity, and are unique
tracks. On the contrary $\tau$ air shower at horizons is spreading its
signal in a wider area  leading to populated (millions-billions) muon and gamma (as well as electron pairs)
  bundles in their showering secondary mode. This amplified  signal may be
observed and disentangled from farer and filtered UHECR, in different ways and places: mountains,
balloons, satellites with different detector array area and thresholds.
The advantage to be in high quota is to be extending the visible target terrestrial
area and solid angle, as well as to let a longer tau flight distance (and energy), and
to enlarge the air shower area; to make an intuitive estimate
the Tau air-shower size area, at tens PeVs-EeVs ,( detectable at
horizons within a lateral  distance as large as $3$ km. from the main
shower axis by a telescope like Magic),is nearly $30 km^2$; at EeV energy the
equivalent detection depth crossed by the tau lepton before the
exit from the Earth reaches $10-20 km$ distances; the corresponding  detection
Neutrino volume (inside the narrow, conic $10^{-3} sr.$, shower beam)
   is within $30-60 km^3$, in any given direction , see \cite{Fargion2005}.
   A few  events of GRBs  a year may be located within these horizons,
   as well as AGN and BL Lac in their flare activity. In such occasions
   Magic, Veritas and Hess array are the most sensitive neutrino telescope at PeVs-EeV energy.
  Even on average, for a present $2\cdot2^o$
 view of Magic, at present energy thresholds, such telescopes
 (for Neutrino at Glashow PeVs energy  windows), are testing
a total mass-solid angle a comparable or larger than to  $10^{-2}  km^3 sr$,
 an  order of magnitude comparable with the present AMANDA detector.
  The very possible existence of blazing Cherenkov noises in present
  Auger, Hires, Magic views should become a very radical signal in such a
  Copernican Tau attitude. The role of an Array in mountains,
  balloons, planes and satellites for horizontal air-showers
  will rejuvenate CR and it will open high energy  Neutrino eyes to the Universe.
     In conclusion  a maximal alert for the Neutrino
     air-showering within the Earth shadows is needed: in AUGER, Milagro,
     Argo, as well as in ASHRA, CRTNT, Shalon Telescopes the signal is
     beyond the corner. In particular   the  Magic (and Veritas) arrays telescopes facing
      from the mountains the Horizons   edges may soon  test our proposal leading to such crown arrays.
      In a sentence we believe that the UHE Neutrino Astronomy
     is beyond the corner, Tau is its courier  and its sky lay just  beneath
     our own  sky: the Earth.

\end{document}